\documentclass[sigconf,9pt,noacm]{acmart}
\renewcommand\footnotetextcopyrightpermission[1]{} 
\renewcommand{\acmConference}{}
\renewcommand{\shortauthors}{}

\usepackage{tikz}
\usepackage{amsmath}

\usepackage{filecontents}
\usepackage{amsbsy}
\usepackage{algorithmic}
\usepackage{balance}
\usepackage{caption}
\usepackage{gensymb}
\usepackage{graphicx}
\usepackage{multirow}
\usepackage{subfigure}
\usepackage{textcomp}
\usepackage{url}
\usepackage{xcolor}

\usepackage[english]{babel}
\usepackage{blindtext}

\usepackage{tikz}
\usepackage{amsmath,amssymb,amsfonts,mathrsfs}
\usepackage{latexsym}
\usepackage{algorithmic}
\usepackage{graphicx}
\usepackage{xcolor}
\usepackage{subfigure}
\usepackage{gensymb}
\usepackage{mathtools}
\usepackage{url}
\usepackage{balance}
\usepackage{xspace}
\usepackage{multirow}
\usepackage{tabularx}
\usepackage{makecell}
\usepackage{textcomp}
\usepackage{caption}
\usepackage{enumitem}
\usepackage{booktabs}
\usepackage{pifont}
\usepackage[vlined,ruled,linesnumbered]{algorithm2e}
\usepackage{pdfpages}
\usepackage{setspace}

\long\def\comment#1{}
\long\def\delete#1{}

\newcommand{\sysname}{\textsc{Matrix}\xspace}

\newcommand{\yijie}[1]{{\color{black}{}#1}}

\newcommand{\sssec}[1]{\vspace*{0.05in}\noindent\textbf{#1} }

\newenvironment{packeditemize}{\begin{list}{$\bullet$}{\setlength{\itemsep}{0.5pt}\addtolength{\labelwidth}{-1pt}\setlength{\leftmargin}{\labelwidth}\setlength{\listparindent}{\parindent}\setlength{\parsep}{1pt}\setlength{\topsep}{0pt}}}{\end{list}}

\begin{document}



\title{A Single-Chain Backscatter Tag for Multi-Sensor Multiplexing}

\author{Yijie Li}
\affiliation{%
	\institution{National University of Singapore}
	\country{}
}
\email{yijieli@nus.edu.sg}

\author{Weichong Ling}
\affiliation{%
	\institution{National University of Singapore}
	\country{}
}
\email{vincentling3gz@gmail.com}

\author{Taiting Lu}
\affiliation{%
	\institution{Pennsylvania State University}
	\country{}
}
\email{txl5518@psu.edu}

\author{Bao Dao}
\affiliation{%
	\institution{University of Massachusetts Amherst}
	\country{}
}
\email{bdao@umass.edu}

\author{Yi-Chao Chen}
\affiliation{%
	\institution{Shanghai Jiao Tong University}
	\country{}
}
\email{yichao0319@gmail.com}

\author{Vaishnavi Ranganathan}
\affiliation{%
	\institution{Microsoft Research}
	\country{}
}
\email{vnattar@microsoft.com}

\author{Lili Qiu}
\affiliation{%
	\institution{UT Austin}
	\country{}
}
\email{lili.qiu.cs@gmail.com}

\author{Jingxian Wang}
\affiliation{%
	\institution{National University of Singapore}
	\country{}
}
\email{wang@nus.edu.sg}


\begin{abstract}
Many real‑world sensing tasks require co‑located, multi‑modal measurements at a single site, typically a bundle of two to five sensors, for example, in plant stress sensing and blood pressure estimation. RF-backscatter devices have emerged as a low-power solution for sensing, yet existing backscatter tags support a single sensor. Placing several single-sensor tags at one site increases attachment footprint and induces mutual coupling between nearby tag antennas, thereby limiting practical deployment. 

We present \sysname, a single-chain multi-sensor backscatter tag that concurrently supports multiple onboard sensors and multiplexes them as a composite voltage, then backscatters it through one analog modulation chain. Rather than time‑division polling, which introduces inter‑sensor sampling offsets, or frequency‑division, which requires independent per‑sensor modulation chains, \sysname  introduces a voltage-division multiplexing architecture in which each sensor value is encoded as a PWM waveform, carrying the measurement in its duty cycle and reserving the amplitude for multiplexing. To support reliable demultiplexing, \sysname selects the voltage‑division weights in a binary‑weighted geometric progression so that every active‑sensor set maps to a uniquely invertible, well‑spaced composite voltage. The composite voltage is then converted into  backscatter frequency shifts through a single modulation chain.  At the receiver, \sysname formulates demultiplexing as a Hidden Markov Model to recover per‑sensor readings while tolerating analog hardware imperfections and multipath. \sysname's ASIC design consumes $25.56\mu W$. Detailed evaluation shows that the prototype, multiplexing five sensors, achieves 20 $dB$ average signal reconstruction SNR at a 30 $kHz$ sampling frequency; we further validate \sysname with case studies in plant sensing, health monitoring, and microphone‑based direction finding.

\end{abstract}
\maketitle

\section{Introduction}
Real-world sensing applications often benefit from co-located, multi-modal sensing: a small bundle of sensors  placed at the same site. On a single leaf,  measurements of leaf temperature, leaf‑surface humidity, and incident light are combined to infer plant stress~\cite{lee2023abaxial,ang2024decoding}. Similarly, in cuffless blood pressure estimation, for human health monitoring, a wearable patch co-locates optical (PPG) and electrical (ECG) sensors, and often integrates motion and skin-temperature sensors within the same small footprint to mitigate noise and compensate for vasomotor variations~\cite{laput2017synthetic,shriram2010continuous}. More broadly, deployments typically instrument one site with two to five co-located sensors spanning multiple modalities~\cite{lee2023abaxial,ang2024decoding,yin2024plant,lu2020multimodal,dang2022iotree,wang2024soilcares,laput2017synthetic}.

RF-based backscatter tags have emerged as a low-power interface for sensing applications by backscattering sensed data onto ambient wireless signals.  However, existing backscatter tags~\cite{ranganathan2018rf,wang2022ultra,sample2008design,sample2008design,na2023leggiero,zhang2016hitchhike,zhang2017freerider,bhat2024zensetag,abedi2020witag} carry only a single sensor, limiting each tag to one sensing modality (e.g., temperature~\cite{pradhan20rtsense} or muscle stretch~\cite{wang20rfidtattoo}). One solution is to place several co-located single-sensor tags, but this increases attachment and physical footprint (e.g., on-leaf, or on-skin) and creates tight spatial proximity that causes mutual coupling~\cite{jin2018wish,jin2018rf}: nearby tag antennas interact through their near fields and perturb each tag's backscattered signal, which severely reduces signal quality. In contrast, devices that support multiple onboard sensors today are not backscatter‑based because they rely on a power-hungry digital processing chain in which a microcontroller (MCU) manages analog-to-digital (ADCs) and digital-to-analog (DACs). \yijie{This raises a question: \textit{Can a single backscatter tag support streaming from multiple sensors?}}

\begin{figure}[t]
	\centering
	\includegraphics[width=1\linewidth]{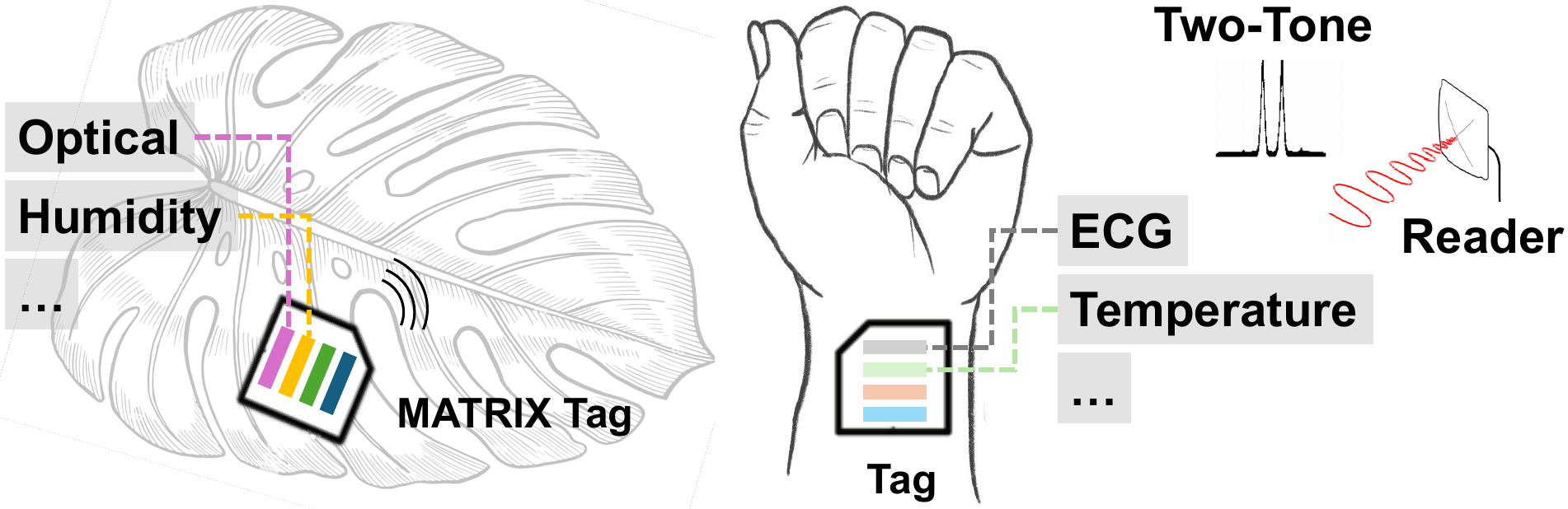}
	\caption{The \sysname tag concurrently acquires and multiplexes multiple onboard sensors and backscatters them via a single modulation chain.} 
	\label{fig:fig1}
	\vspace{-10pt}
\end{figure}

In this paper, we present \sysname, a backscatter tag that concurrently acquires readings from multiple onboard sensors and multiplexes them for backscatter. 
Unlike traditional multiplexing schemes—such as frequency-division multiplexing (FDM), which requires an independent modulation chain per sensor on a single tag or time-division multiplexing (TDM), which incurs unavoidable inter-sensor sampling offsets—\sysname introduces a novel voltage-division multiplexing architecture that concurrently encodes all sensors’ outputs into one superimposed signal and backscatters it as frequency shifts \textit{through a single analog backscatter chain}. Specifically, \sysname first converts each sensor’s output into a pulse-width-modulated (PWM)  waveform on a shared time base, turning time-varying analog readings into duty cycles while keeping acquisition across sensors naturally synchronous. \sysname then carefully selects the voltage-division weights so that the combined signals remain uniquely invertible while fully exploiting the tag’s limited operating voltage range.
At the receiver, \sysname develops a Hidden Markov Model (HMM)-based demultiplexer to reconstruct each sensor stream while remaining robust to analog hardware imperfections. \sysname operates in the $915MHz$ ISM band, consumes 25.56 $\mu W$, and integrates five sensor channels. Detailed experiments show that \sysname achieves 20 dB average SNR in signal reconstruction\footnote{The log-ratio of ground-truth signal power to reconstruction-error power.} at $30kHz$ sampling frequency with five sensors, meeting the needs of a wide range of sensing applications (see Fig.~\ref{fig:fig1}); we further validate \sysname with case studies in wearable health monitoring, plant sensing, as well as acoustic based direction finding using microphones which requires high-frequency (a few kHz) sampling.

\sysname's main objective is to  multiplex readings from multiple onboard sensors onto a backscattered signal.
A strawman approach, time-division multiplexing (TDM), sequentially polls sensors with a local clock, introducing inter-sensor sampling offsets. Frequency-division multiplexing (FDM) acquires sensors concurrently by assigning each sensor a distinct frequency band, but requires an independent modulation chain per sensor with its own  voltage‑controlled oscillator (VCO) and mixers, increasing hardware complexity.  In contrast, \sysname  introduces a  novel \textbf{voltage-division multiplexing (VDM)} architecture that superimposes concurrent sensor readings into one composite voltage signal, which is backscattered through a single analog modulation chain. With VDM, each sensor is assigned a distinct multiplexing weight so their sum can be inverted (demultiplexed) to recover individual readings. However, directly weighting and summing raw sensor readings is non‑invertible: different sets of readings can produce the same weighted sum. To address this, \sysname represents each sensor reading in the time domain: it encodes each sensor value as a PWM waveform on a shared time base, while reserving amplitude for multiplexing via distinct voltage weights. Thus at any instant in a PWM cycle, a sensor contributes either zero or its assigned weight, \yijie{producing discrete composite levels tied to active sensors} (where ``active'' means PWM high at that instant).  \yijie{The shared time base naturally ensures synchronous acquisition across sensors}. 

\yijie{With each sensor's reading encoded in its PWM duty cycle and the PWM amplitude reserved for multiplexing, \sysname's multiplexing reduces to a weight-selection problem. The design challenge lies in choosing weights such that every sensor combination maps to a unique voltage level within the tag’s range (e.g., up to 1V). Prime-based weights guarantee uniqueness but grow superlinearly, forcing rescaling and reducing frequency level spacing, which hurts robustness in decoding. In contrast, \sysname  uses binary-weighted progression for uniform spacing, efficiently using voltage range and improving demultiplexing robustness. Thus, \sysname's VDM combines concurrent readings into a uniquely invertible weighted sum, which is then backscattered using only one analog modulation chain, cutting hardware overhead while preserving signal integrity. Encoding sensor streams as PWM requires a periodic sawtooth time base; Sec.~\ref{sec:clock_free} details how \sysname derives this timing directly from ambient RF without an onboard clock. }

At the receiver, the composite voltage appears as a backscattered signal whose instantaneous frequency steps \yijie{reflect active sensor combinations}. The receiver therefore must demultiplex this signal to reconstruct the individual sensor streams by determining which frequency level is present and how long it persists.  \yijie{A naive approach like Short-Time Fourier Transform (STFT) suffers from resolution trade-offs, blurring short frequency levels and spreads transition instants.} To address this, \sysname designs a phase-difference-based instantaneous frequency tracking algorithm to recover the frequency-time trace. The next step is to segment the trace into  PWM cycles. A simple fixed threshold—e.g., marking a new cycle when the frequency returns to the top level—works on clean traces but is sensitive to hardware-induced frequency jitter, causing false or missed starts. \sysname therefore develops a two-stage trace segmentation to reliably locate each PWM cycle. The remaining challenge \sysname must address is to pinpoint, within each cycle,  frequency transition instants despite analog hardware imperfections and multipath. For example, the VCO and RF switch maps the composite voltage to backscattered frequency shifts; however, a change in the composite voltage level does not always produce a clean, instantaneous jump in backscattered frequency and can instead glide, creating 
ambiguous frequency transitions. Merged transitions caused by identical sensor readings further complicate demultiplexing.  \sysname resolves this by formulating a probabilistic decoder. Specifically, \sysname builds a Hidden Markov Model (HMM) over the finite set of sensor-combination states to infer the most likely state sequence and their durations. This design enables reliable multi-sensor reconstruction even when PWM intervals merge due to identical sensor readings and despite analog hardware imperfections and multipath.

We implement \sysname's prototype on a two-layer PCB with commercial off-the-shelf  components. The prototype has a compact form factor (3$cm$ by 2$cm$) and costs $\$17$ per tag, providing five sensor channels. The ASIC design of \sysname consumes 19.32 $\mu W$ for  multiplexing and 7.47 $\mu W$ for timing extraction. The two-layer PCB prototype draws power between  $1.36 mW \sim 2.32 mW$, depending on the number of sensors (1—5). \sysname operates in the 915 MHz band, driven by a two-tone excitation signal from a USRP N210.  We evaluate \sysname's performance with 10 different sensor types and test across diverse applications. Experiments are conducted in both line-of-sight and non-line-of-sight scenarios, as well as in environments with moving objects. Our results show that:

%

\begin{packeditemize}
	
	\item \textbf{Overall Capability:} \sysname achieves an average SNR of $20~dB$ in signal reconstruction, and supports 
    five sensors with a sampling frequency of 30 $kHz$.

    \item \textbf{Multimodal Sensing:} \sysname tag, deployed on a single leaf, reconstructs four sensors with 44 dB average SNR; deployed as a wearable patch, it tracks co-located 3-axis accelerometer, PPG and ECG sensors with 47 dB average SNR.
	
	\item \textbf{High-frequency Acoustic Sensing:}  With four microphones on the tag, \sysname achieves acoustic angle-of-arrival (AoA)  errors of $4\degree$ for a 1 $kHz$ input and $6\degree$ for a 3 $kHz$ input, outperforming the TDM baseline by $10\degree$ across tested angles. 
	
\end{packeditemize}

\sssec{Contributions: }\sysname's core contributions include:

\begin{packeditemize}
	\item A backscatter tag that concurrently acquires and multiplexes   multiple onboard  sensors over one analog modulation chain.
	
	\item A novel voltage-division multiplexing  architecture that encodes sensor readings as PWM and multiplexes them with binary-weighted geometric weights for uniquely invertible, well-spaced composite levels.
	
	
	\item An HMM demultiplexer that robustly reconstructs individual sensor streams despite hardware imperfections and multipath.
	
	\item Extensive evaluation with various case studies including plant sensing, health monitoring, and acoustic AoA estimation.
\end{packeditemize}

\section{Related work}


\sssec{RFID-based Sensing:} 
RFID tags~\cite{sanpechuda2008review} have been widely studied   for object identification~\cite{ilie2011survey,su2020partitioning,roberts2006radio,yang2024rf,yin2024efficient}, localization~\cite{wang13dude, yang14tagoram, chuo17rfecho, wang17dwatch, chang18rfcopybook, wang2022ultra}, and motion tracking applications~\cite{bryce14bringing, shangguan17enabling, gao18livetag, wang20rfidtattoo, ji2024construct}. By leveraging the antenna impedance coupling effects, RFID tags extend  beyond location tracking to detect material  property changes. Recent studies demonstrated  material identification~\cite{wang17tagscan, xie19tagtag}, humidity and temperature sensing~\cite{wang20soil, pradhan20rtsense, POTYRAILO2013587, khan22estimating}, and even more complex applications such as monitoring egg incubation~\cite{sun24rfegg}, fruit ripeness~\cite{karmakar2024metasticker}, gas~\cite{sun2024gastag}, pesticide~\cite{he2024hornbill} by integrating various functional materials~\cite{sun2024gastag,wang2018challenge,wang17tagscan, xie19tagtag,wang2024soilcares,he2024hornbill} into the antenna. However, they are typically designed for task-specific applications, with each implementation tailored to a single sensing modality. In contrast, \sysname is a general-purpose backscatter tag that supports  concurrent sensing from multiple onboard off-the-shelf sensors.

\sssec{Backscatter Sensor Interface:} Researchers have built custom  backscatter tags that interface with off-the-shelf  sensors. WISP~\cite{sample2008design} uses an onboard micro-controller for programmability and an ADC for data acquisition. However, these digital components are power-hungry, and WISP consumes over $1 mW$ even with a single  sensor. While recent studies~\cite{ranganathan2018rf,wang2022ultra,sample2008design,na2023leggiero,bhat2024zensetag,abedi2020witag} reduce tag power by  offloading computation-intensive digital signal processing to the reader, they still support only one sensor per tag.  Deploying several single-sensor tags at one site increases physical footprint and hardware overhead, and tight spacing induces mutual coupling among nearby tag antennas, which is undesirable for applications that require multiple  co-located sensors.  Instead, \sysname's tag concurrently streams multiple onboard sensors and consumes less than $2.32 mW$ while multiplexing 5 sensors, and its ASIC design consumes just $25.56\mu W$.


\sssec{Multi-Sensor Backscatter:} 
Recent work has explored backscattering multiple  sensors from a single tag: A TDM approach \cite{zhao2021microphone} sequentially polls onboard sensors with a $\$500$ FPGA; however, this  introduces inter-sensor sampling offsets: even a 10 $\mu$s offset yields $10.8\degree$ phase error at 3 kHz, degrading performance in time-sensitive tasks like acoustic AoA estimation~\cite{yun2017strata,wang2019millisonic,ding2020rf,li2022experience}. Na\"{\i}ve post processing is insufficient  to compensate these errors, as offsets further wander with clock drift  due to crystal oscillator instability and temperature variations~\cite{TIdrift,ADdrift,TIdrift2}, especially with inexpensive clock, making them unpredictable over time. While actively synchronizing the tag clock to the external infrastructure temporarily reduces drift, it adds power and hardware complexity on the tag. As a result, prior multi-sensor tags that use TDM~\cite{zhao2021microphone} pay for precise sequential polling with a costly FPGA priced over $\$500$~\cite{FPGAcost}. By contrast,   backscatter designs that achieve  concurrent  acquisition~\cite{mandal2010lowpowerbodysensor} are limited to spike-like physiological signals (e.g., PCG and PPG). Finally, while frequency-division multiplexing could, in principle, enable concurrent sensing by assigning distinct frequency bands to sensors~\cite{ranganathan2018rf},  it requires separate modulation chains per sensor, leading to substantial power and footprint overhead. In contrast, \sysname is a \textit{single-chain backscatter tag that concurrently samples multiple onboard sensors.} It multiplexes all sensor readings as a composite voltage, and backscatters it through one analog modulation chain. This design reduces the physical footprint and avoids inter-sensor  offsets;  Sec.~\ref{sec:application} shows that \sysname outperforms a TDM baseline by $10\degree$ in AoA estimation.

\section{Overview}
\begin{figure}[t]
	\centering
	\includegraphics[width=1\linewidth]{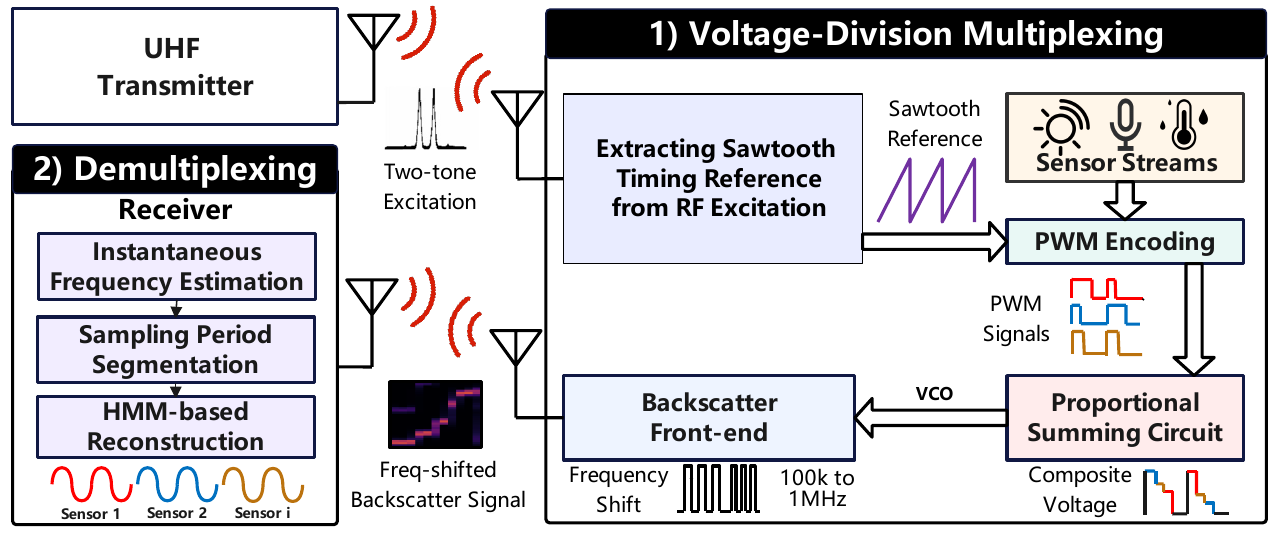}
	\vspace{-20pt}\caption{System overview of \sysname.}
	\label{fig:overview}
	\vspace{-10pt}
\end{figure}

\sysname is a single-chain backscatter tag that concurrently acquires multiple onboard sensors, multiplexes them into one voltage using a novel VDM scheme, and backscatters through one analog modulation chain. \sysname achieves this by (1) encoding each sensor value as a PWM duty cycle on a shared time base, dedicating the PWM amplitude to multiplexing;  (2) choosing voltage-division weights so that, in each PWM cycle, every combination of sensors that are PWM high sums to a distinct composite voltage level, while the resulting levels are well separated within the tag’s operating-voltage range; and  (3) at the receiver, modeling demultiplexing as a Hidden Markov Model that robustly recovers per‑sensor readings despite analog hardware imperfections and multipath.

Fig.~\ref{fig:overview} summarizes  \sysname's pipeline. On the tag, each sensor value is compared against a shared sawtooth timing reference (derived from an ambient RF carrier) to produce a PWM waveform; since all PWMs share the same time base, acquisition is naturally synchronous.  Each  PWM is assigned a distinct weight, and the weighted PWMs are summed into a single composite voltage that is  invertible to the per-sensor contributions within each cycle. A low-power VCO converts this voltage to an output frequency, and an RF switch mixes it with the ambient carrier, producing discrete backscatter frequency levels that change within a PWM cycle according to which sensors are active (PWM high) at each instant. On the receiver side, \sysname estimates the frequency over time, segments it into PWM cycles, and employs an HMM based demultiplexer to infer the sequence of  frequency levels and their level durations; these durations give per-sensor duty cycles, which are then mapped back to sensor readings. The rest of the paper addresses the following technical challenges:

\begin{figure}[t]
	\centering
	\includegraphics[width=0.78\linewidth]{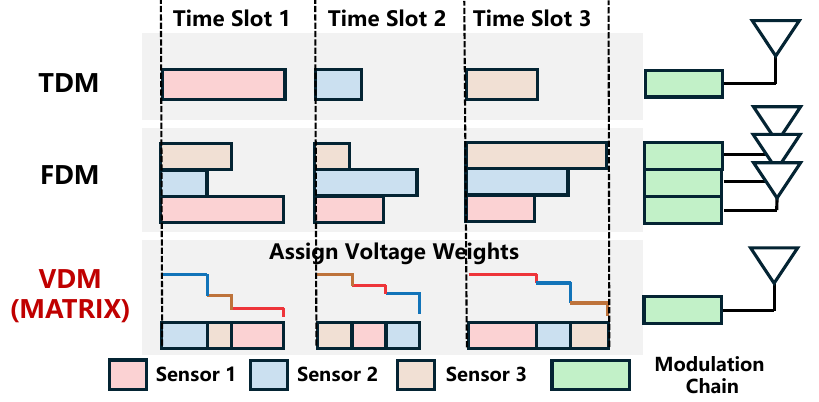}
	\vspace{-10pt}
	\caption{TDM introduces inter-sensor sampling offsets. FDM requires multiple independent modulation chains for each sensor. In contrast, \sysname proposes a voltage-devision multiplexing (VDM) to sum all sensor readings and backscatter through a single modulation chain. } 
	\label{fig:vdm_compare}
	\vspace{-15pt}
\end{figure}
 
\sssec{(1) Voltage-Division Multiplexing:} The key challenge in VDM is to let  many sensors share one analog modulation chain while keeping the summed signal invertible. Simply weighting and summing continuous sensor values is not invertible: different combinations can produce the same total. \sysname addresses this by encoding each sensor value as a PWM duty cycle and dedicating the PWM amplitude to multiplexing via distinct per‑sensor voltage weights; the weighted sum then takes values from a small discrete set of voltage levels determined by the active sensors. A second challenge is choosing weights so that all summed voltage levels stay within the VCO input range. Prime‑based weights ensure uniqueness, but their subset sums grow super-linearly and create highly uneven spacing among individual voltage levels; fitting them into the VCO’s range requires rescaling that squeezes the uneven inter‑level spacing and reduces robustness to noise. \sysname instead uses  a binary-weighted geometric progression, creating uniform inter-level spacing and distinct  voltage levels for every sensor combination, supporting robust demultiplexing. Sec.~\ref{sec:sync} details our approach.

\sssec{(2) Demultiplexing:} The challenge at the receiver is to recover, within each PWM cycle, the sequence of discrete  backscatter frequency levels and their durations. Windowed spectral methods (e.g., STFT) face an inherent time–frequency tradeoff: short frequency levels smear in long windows, and adjacent  levels blur in short windows. \sysname instead estimates the frequency-time trace from adjacent-sample phase differences. Another challenge is segmenting PWM‑cycle boundaries in the presence of frequency jitter; \sysname develops a two‑stage segmentation to locate periods reliably. Finally, pinpointing level transitions inside each cycle is challenging under analog hardware imperfections and multipath. \sysname casts  transition detection as an HMM and applies the Viterbi algorithm to infer the sequence of frequency  levels and their durations, which determine per‑sensor PWM duty cycles and thus sensor readings. Sec.~\ref{sec:demodulation} details our approach.

\section{Multi-Sensor Multiplexing}\label{sec:sync}

This section describes how \sysname multiplexes multiple sensor streams concurrently and backscatters them using only one analog modulation chain. Traditional TDM unavoidably introduces inter-sensor sampling offsets.  Frequency-division multiplexing, where each sensor  transmits over a distinct frequency band, can acquire sensors concurrently, but  requires multiple independent modulation circuits on a single tag. In contrast, \sysname introduces a minimalist approach—it sums per-sensor readings   into one composite voltage  and backscatters it through a modulation chain consisting of just one voltage-controlled oscillators (VCO) and RF switch (shown as Fig.~\ref{fig:vdm_compare}). A key challenge is ensuring that, even after summation, individual sensor readings can be   uniquely demultiplexed. Rather than na\"{\i}ve summation, which can obscure sensor identity, \sysname builds a novel voltage-division multiplexing (VDM) architecture that ensures the composite maps uniquely to the original streams.

Specifically, \sysname: (1) encodes each sensor’s value as a pulse-width modulation (PWM) duty cycle on a shared time base—reserving the PWM amplitude for multiplexing (Sec.\ref{sec:parallel}); and (2) carefully selects per-sensor voltage weights so that every combination of active sensors maps to a distinct, invertible composite level with comfortable spacing within the tag’s operating-voltage range (Sec.\ref{sec:multiplex}).

\subsection{PWM Encoding of Raw Sensor Readings}\label{sec:parallel}
The role of \sysname's VDM is to enable multiple sensors to share one analog chain for backscatter: each sensor is assigned a distinct voltage weight and their sum is backscattered through one analog modulation chain. A key requirement is that  the sum must  be uniquely invertible, i.e., the mapping from per‑sensor values to the composite must be one‑to‑one. However, applying weights directly to raw sensor readings yields a non‑invertible mapping: different sets of per-sensor values can produce the same weighted sum, so the individual signals cannot be uniquely demultiplexed.

\begin{figure}[t]
	\centering
	\includegraphics[width=0.95\linewidth]{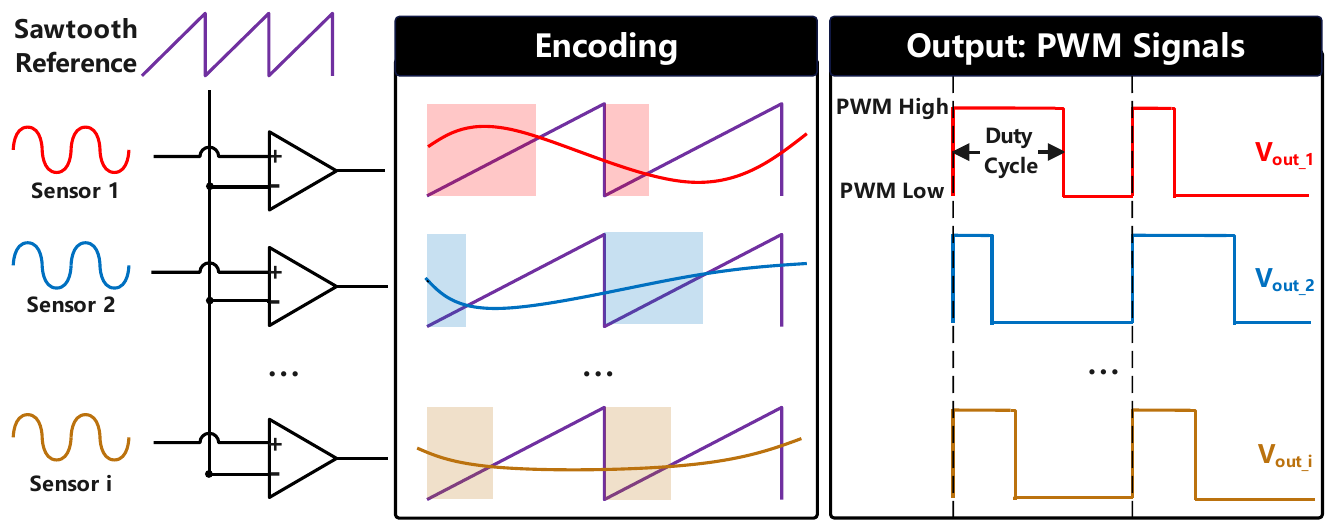}
	\vspace{-9pt}
	\caption{\sysname encodes readings from multiple sensors into  PWM signals on a shared time base (PWM's amplitude reserved for multiplexing); parallel comparators compare each input to a shared sawtooth timing reference, synchronizing sampling across sensors.} 
	\label{fig:pes}
	\vspace{-15pt}
\end{figure}

To resolve this ambiguity, \sysname deliberately encodes all sensor inputs as PWM signals on a shared time base (Fig.~\ref{fig:pes}):  the duty cycle (PWM-high durations) carries the measurement, while the PWM amplitude is set to a distinct voltage weight for multiplexing. 
With PWM encoding, when a sensor’s PWM is high (defined as an active sensor), it contributes its assigned voltage weight, and when it is low (inactive sensor), it contributes zero; thus, the  sum takes only a  set of discrete levels corresponding to the active sensors. Since all sensor signals are encoded as PWM with a shared timing reference, acquisition across sensors is naturally synchronous. 

\begin{figure*}[t]
	\centering
	\includegraphics[width=0.9\linewidth]{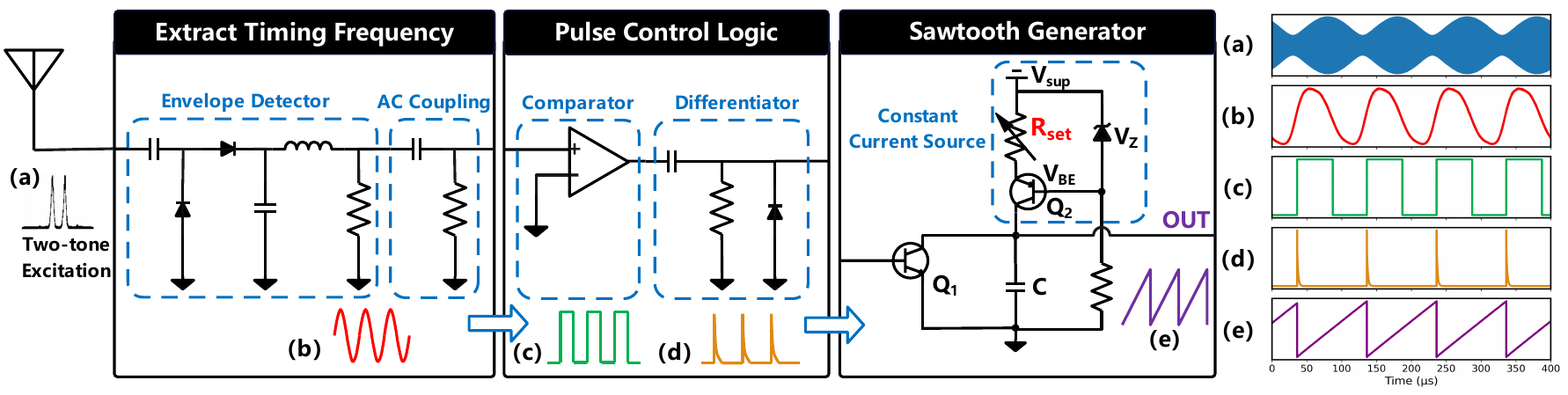}
	\vspace{-10pt}
	\caption{Circuit flow of \sysname's sawtooth-timing extraction from ambient signals: (a) ambient two-tone RF carrier;  (b) envelope extraction yielding the timing frequency; (c) amplitude-invariant square wave via a ground-referenced comparator;  (d) RC differentiator and diode generating narrow timing pulses; (e) sawtooth generation via a constant-current source; a tunable $R_{\text{set}}$ adapts to on-the-fly sampling-rate tuning.} 
	\label{fig:env_detect}
	\vspace{-10pt}
\end{figure*}



\sssec{PWM Encoding with Comparators:} To encode the sensor reading as PWM, 
\sysname compares the sensor reading  against a sawtooth timing reference using a $\mu W$ comparator (TLV3201~\cite{TLV3201}). When the sensor value exceeds the voltage of the rising sawtooth, it outputs a high voltage; otherwise, it remains low (see details in Fig.~\ref{fig:pes}). This linear rising slope allows the comparator to generate a PWM signal where the duty cycle is proportional to the sensor's analog voltage,  effectively encoding the original sensor readings. Specifically, assuming the input signal's amplitude of the $i^{th}$ sensor stream is $V_{i}$, and the sawtooth waveform has a maximum amplitude $V_{max}$ and frequency $f_s$, the resulting PWM duration $T_{d_i}$ is given by: 


\begin{equation}
\label{equa:recover}
T_{d_i} = \frac{V_{i}}{V_{max}}\cdot\frac{1}{f_s} + \epsilon
\end{equation}
where $\epsilon$ is a  constant delay (less than 2 ns jitter) introduced by the comparator~\cite{TLV3201}. Generated PWM signals vary in duty cycles (determined by the sensor reading) but share the same voltage level $V_s$ prior to weight assignment, bounded by the  tag’s maximum operating-voltage range. In Sec.~\ref{sec:multiplex}, we assign each sensor a distinct weight relative to $V_s$ for multiplexing.



\sssec{Concurrent Acquisition: }\sysname samples all sensors concurrently:  parallel  comparators compare each sensor reading against the same sawtooth timing reference (Fig.~\ref{fig:pes}).  Sharing one time base makes the resulting PWM outputs naturally synchronous—no inter-sensor offsets and no scheduling delays—unlike traditional TDM, in which sensors are sampled sequentially by a local clock. The only residual skew comes from comparator propagation delay ($<2$ ns)~\cite{TLV3201}, which we verified with an oscilloscope.

\begin{figure}[t]
	\centering
	\includegraphics[width=1\linewidth]{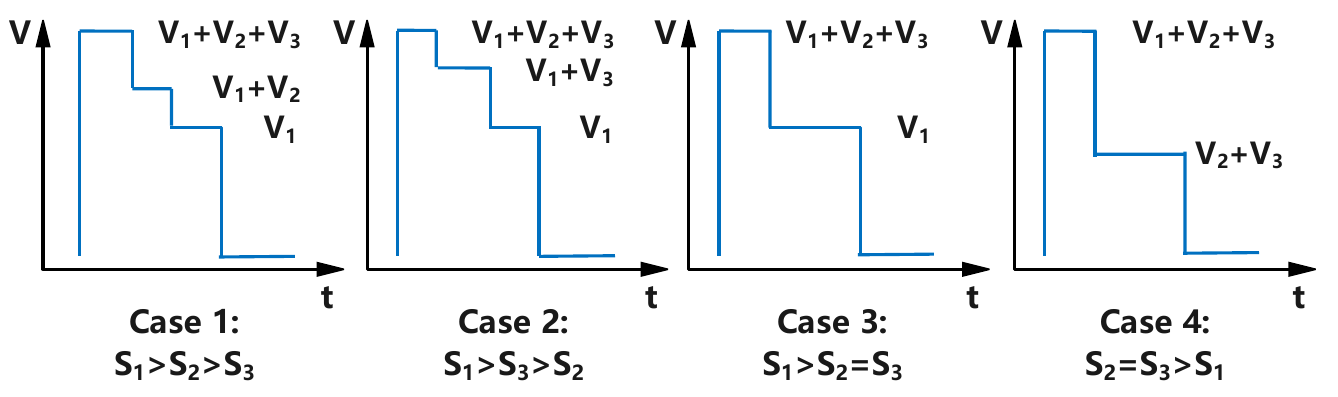}
	\vspace{-20pt}
	\caption{Example of three-sensor voltage-division multiplexing over one PWM cycle: $S_i$ are sensor values (duty cycles) and $V_i$ their voltage weights. At any instant, each sensor contributes 0 or its weight, so the sum can take only one of the eight subset sums of $\{V_1,V_2,V_3\}$.} 
	\label{fig:binary}
	\vspace{-10pt}
\end{figure}

\subsection{RF-derived Sawtooth Timing}\label{sec:clock_free} 

\sssec{Extracting Timing from an Ambient Two-Tone Carrier:} As Sec~\ref{sec:parallel} describes, encoding sensor values as PWM requires comparing them against a shared sawtooth timing reference, which sets the sampling rate for data acquisition across all onboard sensors. While this reference is typically generated by a local clock, \sysname derives it directly from an ambient two-tone RF carrier. Specifically, if the excitation source emits tones at $\omega_1$  and $\omega_2$ as shown in (Fig.\ref{fig:env_detect} (a)), the envelope of the received two-tone signal  oscillates at the difference angular frequency $\omega_{\mathrm{env}} = |\omega_1-\omega_2|$. For example, tones at 915 MHz and 915.01 MHz yield an envelope frequency of 10 kHz; adjusting the tone spacing tunes the tag’s  sampling rate on the fly. To obtain the envelope, \sysname uses a cascaded rectifier (HSMS285C) followed by a low-pass filter (Fig.\ref{fig:env_detect} (b)).


\sssec{Accounting for Multipath and Sawtooth Generation:} The amplitude of the two-tone envelope can fluctuate with multipath,  destabilizing the timing reference. To mitigate this, \sysname couples the envelope  through a series of capacitors to remove the DC offset, then feeds it to a ground-referenced comparator that produces an amplitude-invariant square wave (see Fig.~\ref{fig:env_detect} (c)). This square wave is then passed through an RC differentiator and diode (1N4148) to generate narrow pulses that gate a constant-current  source, producing a sawtooth used as the PWM timing reference (Fig.\ref{fig:env_detect} (d–e)).

\begin{figure}[t]
	\centering
	\vspace{2pt}
	\includegraphics[width=1\linewidth]{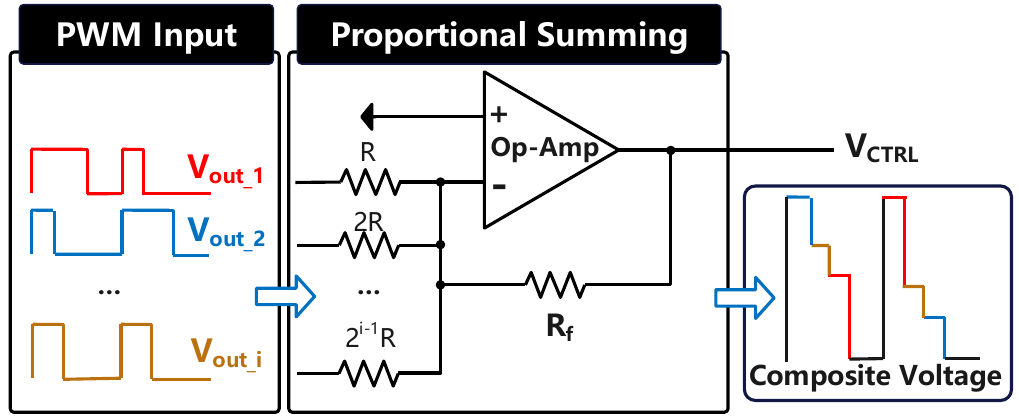}
	\vspace{-20pt}
	\caption{Voltage-division sum of per-sensor PWM signals, producing the composite  voltage.} 
	\label{fig:summing}
	\vspace{-10pt}
\end{figure}

\subsection{Voltage-division Weight Selection}\label{sec:multiplex}

With each sensor value encoded as  PWM duty cycle, and the PWM amplitude reserved for multiplexing, \sysname must choose per-sensor 
voltage weights so that  every  sensor combination maps to a distinct   summed voltage.  Fig.~\ref{fig:binary} shows the sum of three PWM-encoded sensor signals over one PWM cycle:  $S_1, S_2,S_3$ denote their duty cycles, and $V_1,V_2, V_3$ are the  voltage weights assigned to each sensor. Within that cycle, at any instant each  sensor contributes either 0 (PWM low) or its weight (PWM high), so the composite (sum) voltage can take only one of $[V_1,V_2, V_3, V_1+V_2, V_1+V_3, V_2+V_3, V_1+V_2+V_3]$. The design goal is to choose the weights so that the composite can be uniquely inverted back into the individual streams, even when two or more  sensors  report the  same value within a cycle (e.g., Case 3 and Case 4). 

\sssec{Weight Selection: }A strawman approach is to assign distinct prime-valued voltage weights (2, 3, 5, …) to the sensors. While this ensures uniqueness of the weighted sum, decodability alone is insufficient:  the composite voltage levels must also exhibit sufficient inter-level spacing within the tag’s maximum operating-voltage range to support robust demultiplexing at the receiver. However, the subset sum of primes grow super-linearly with the number of sensors, causing widely varying gaps between adjacent levels. To fit all levels within the VCO’s allowable voltage range (e.g., up to 1 V), the entire set must be rescaled, which compresses all gaps proportionally. As a result, the smallest inter-level gaps become exceedingly narrow, making the demultiplexing hard and highly sensitive to noise (e.g., hardware imperfections).


Instead, \sysname assigns the voltage weight following a binary-weighted geometric progression.  This ensures uniform inter‑level spacing, efficiently using the available voltage range and improving reliability.  To do this, each PWM encoded signal  is fed through a dedicated resistor before summation, with values:
\begin{equation}\label{eqn:progression}
	R_i =\alpha R_f\cdot 2^{i-1}
\end{equation}
where $R_i$ is the resistor value assigned to the $i^{th}$ sensor stream, $R_f$ is the feedback resistor, $\alpha$ is a preset coefficient, $i$ represents the sensor stream index. This realizes the voltage weight assignment for each sensor based on the geometric progression.
With this  resistor configuration, the composite voltage $V_{CTRL}$ of all $N$ sensor streams is given by:
\begin{equation}
	V_{CTRL} = R_f \sum_{i=1}^{N} \frac{V_s}{R_i}
\end{equation}
where $V_s$ represents the original voltage of each sensor’s PWM signal, which is the same across all signals. This summation is performed via a  summing circuit consisting of a low-power operational amplifier (AD8605)  as shown in Fig.~\ref{fig:summing},  which superimposes the weighted voltages from multiple sensor streams into a single composite voltage.

\begin{figure}[t]
	\centering
	\includegraphics[width=0.8\linewidth]{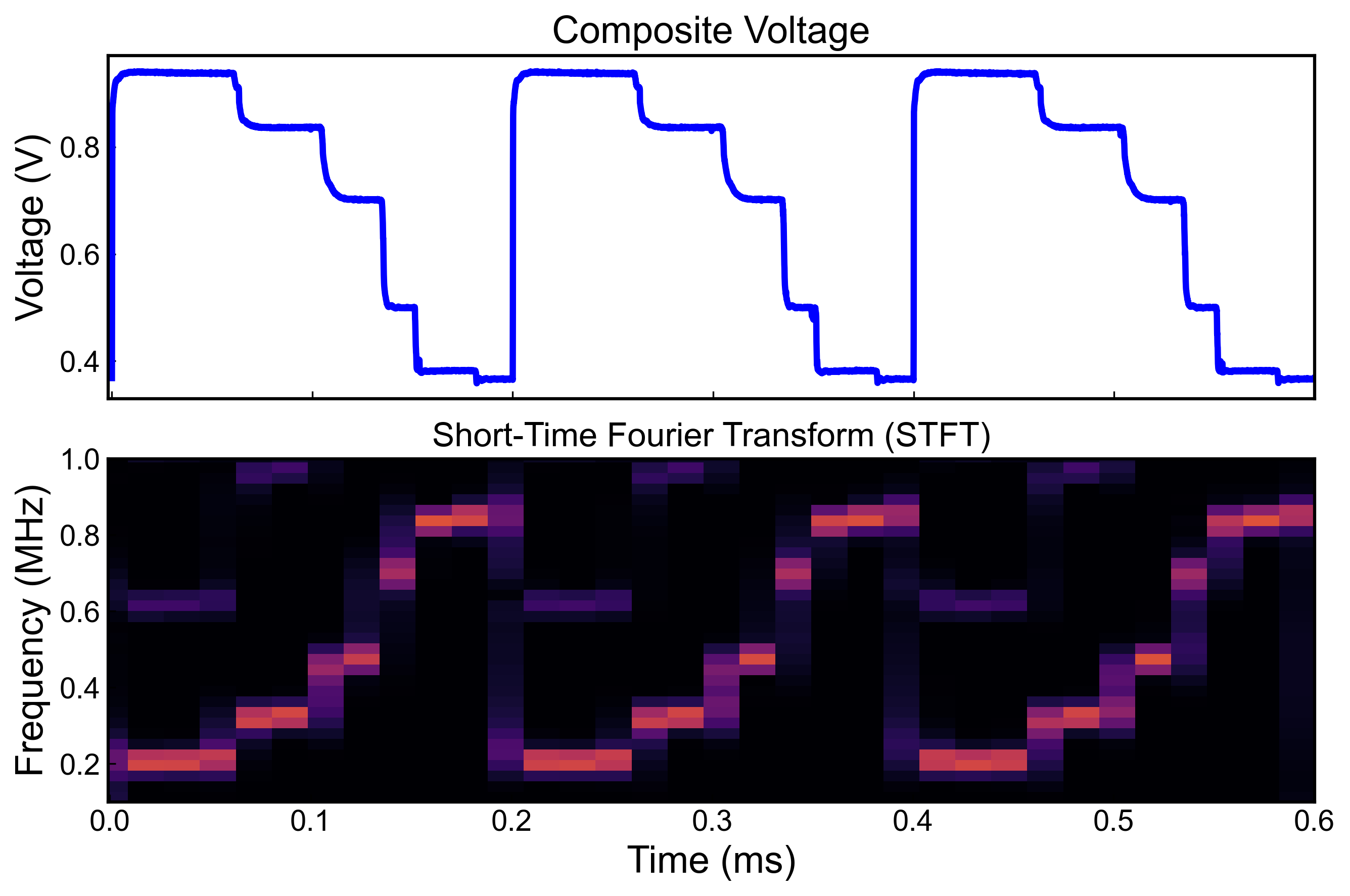}
	\vspace{-10pt}
	\caption{Real-world snapshot of the composite  voltage and corresponding backscatter frequency  for a five-sensor tag over three PWM cycles at a 5 kHz sampling frequency.} 
	\label{fig:v2f}
	\vspace{-10pt}
\end{figure}

The resistor configuration above in Eqn.~\ref{eqn:progression} ensures that every  combination of sensors produces a distinct  composite voltage, enabling the receiver to demultiplex and reconstruct individual sensor readings. Furthermore, the geometric progression of resistor values allows uniform voltage spacing between sensor combinations, improving reliability.

\sssec{Backscatter Over A Single Chain:} Building on the VDM architecture, the  readings from all onboard sensors sum to a single composite voltage at each instant, which is then backscattered via one analog modulation chain based on \cite{ranganathan2018rf}. Specifically, \sysname maps this voltage to an output frequency with a  low-power voltage-controlled oscillator (VCO) LTC6990~\cite{LTC6990}; over the operating range the map is linear and inverse (a higher voltage yields a lower frequency output). 
\sysname then uses an RF switch (ADG902~\cite{ADG902}) to modulate the ambient 915 MHz carrier to produce discrete backscatter frequency shifts determined by the active sensor combination.  In a five-sensor configuration, the 32 possible sensor combinations are realized within a 1 MHz frequency band. Fig.~\ref{fig:v2f} shows a real-world snapshot of the composite voltage ($V_{out}$) and the corresponding backscatter frequency shift ($f_{out}$) for a five-sensor tag over three consecutive PWM cycles at a 5 $kHz$ sampling frequency.

\begin{figure*}
	\centering
	\begin{minipage}{0.31\linewidth}
		\includegraphics[width=\linewidth]{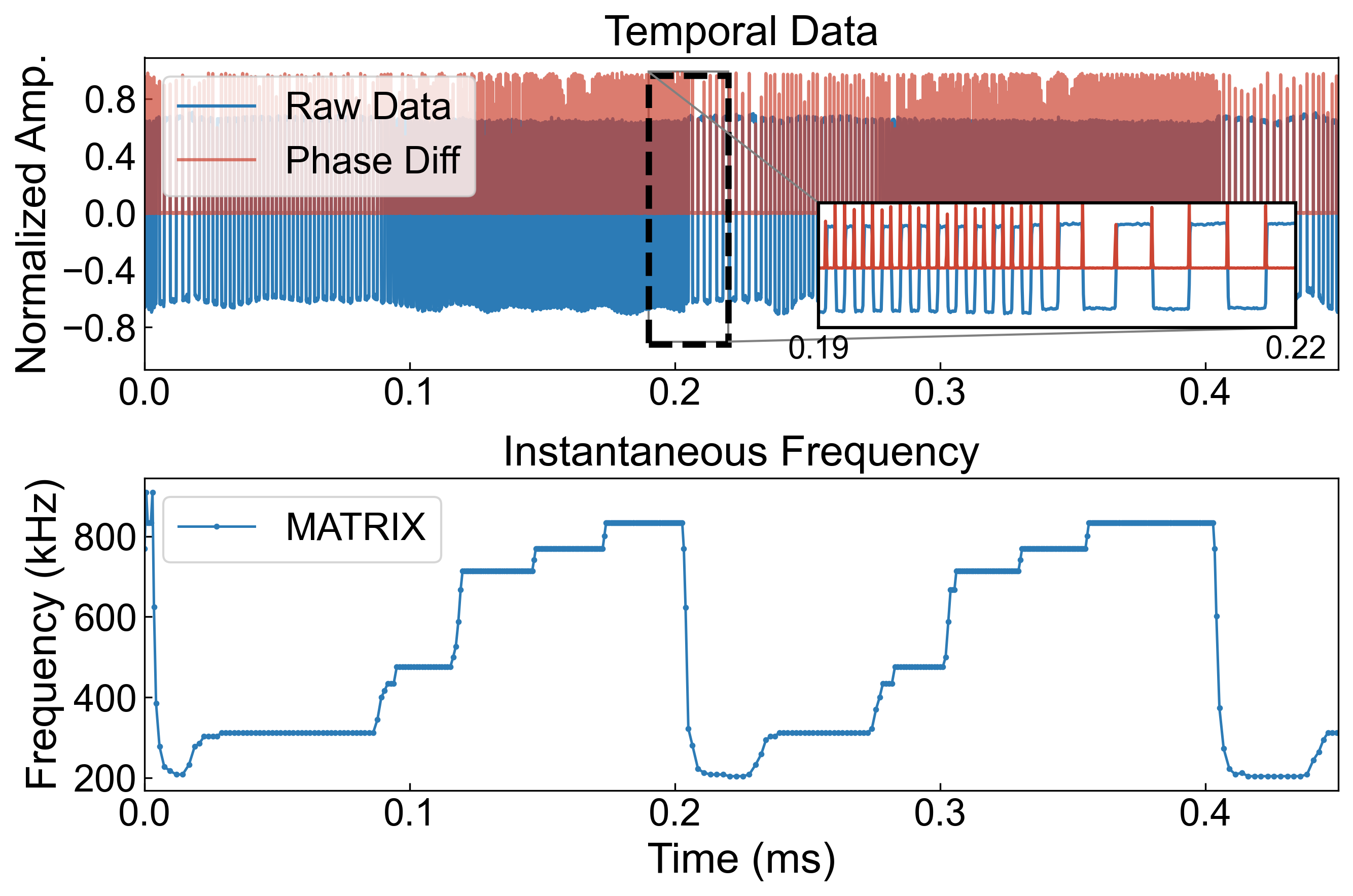}
		\vspace{-15pt}
		\caption{Top: received baseband signal and adjacent sample phase difference. Bottom: estimated  backscatter frequency shifts. } 
		\label{fig:inst_Freq}
	\end{minipage}\hspace{15pt}
	\begin{minipage}{0.31\linewidth}
		\includegraphics[width=\linewidth]{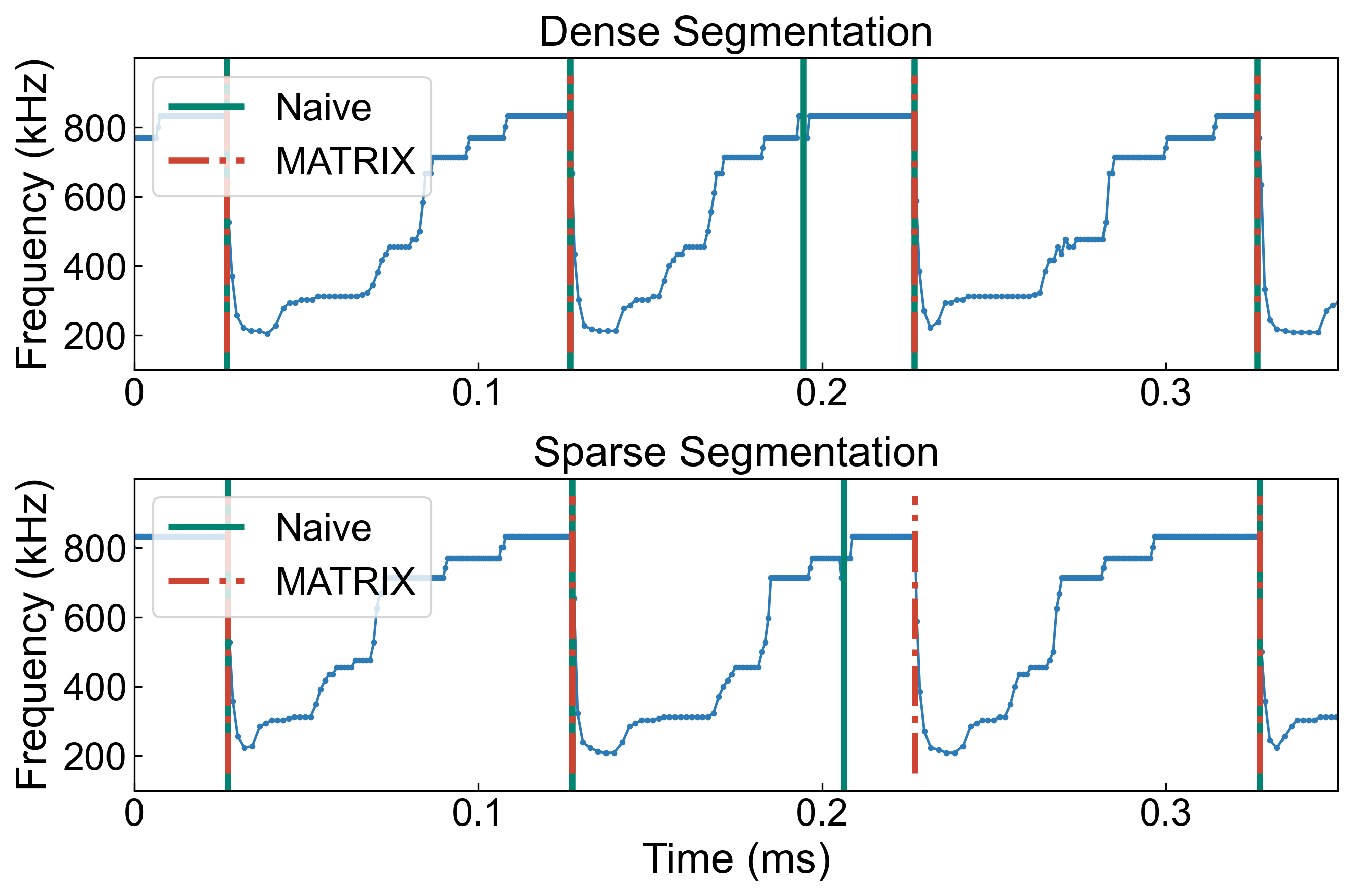}
		\vspace{-15pt}
		\caption{Segmentation of the frequency trace into PWM cycles. A na\"{\i}ve fixed threshold gives too-dense or too-sparse cuts under frequency jitter. } 
		\label{fig:period}
	\end{minipage}\hspace{15pt}
	\begin{minipage}{0.31\linewidth}
		\includegraphics[width=\linewidth]{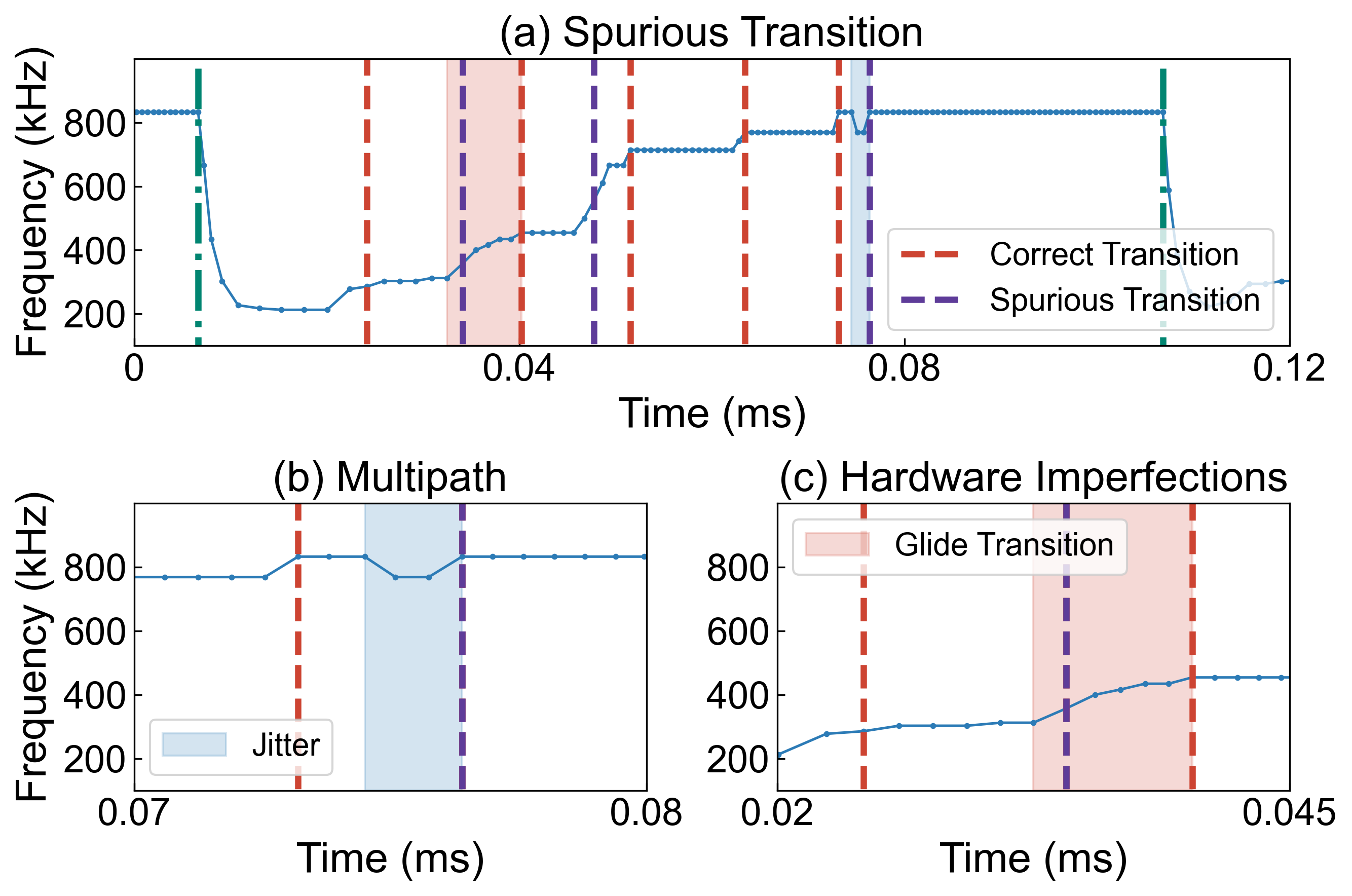}
		\vspace{-15pt}
		\caption{Analog hardware imperfections and multipath create spurious transitions that mislead simple threshold based detectors.} 
		\label{fig:transit}
	\end{minipage}
	\vspace{-10pt}
\end{figure*}

\section{Demultiplexer}\label{sec:demodulation}
\sysname's tag acquires readings from all onboard sensors, sums them into a single composite voltage via VDM, and maps that voltage to backscatter frequency shifts. At the receiver, \sysname's demultiplexing  has three stages. 1) \sysname must track the backscatter frequency over time; at any moment, the frequency takes one of discrete levels, each uniquely associated with a  combination of active sensors.  2) \sysname must segment this frequency trace into PWM cycles,  each cycle representing a multiplexed sample of all sensors. 3) Lastly, within each cycle \sysname must pinpoint the frequency transition  instants between levels and measure how long each level persists; these level durations are the key to recovering the precise per-sensor readings.

\subsection{Frequency Trace Estimation and Segmentation}\label{subsec:freq_esti}
The reader must first recover a frequency‑versus‑time trace and then segment it into PWM cycles before inferring per‑sensor values within each cycle.

\sssec{Why not STFT?: }Traditional window-based spectrum analysis methods such as Short-Time Fourier Transform (STFT) create a time-frequency resolution trade-off that blurs short level durations and smears transition instants. For example, a long analysis window improves frequency resolution but assumes the frequency remains stationary within the window, making it difficult to resolve short-duration frequency levels. Conversely, a short window can better localize rapid frequency transitions in time but yields coarse frequency resolution, causing closely spaced frequency levels to blur or merge. Such a trade-off violates the demultiplexing requirement for simultaneously achieving high time and frequency resolution. In Fig.~\ref{fig:v2f}, the top panel shows the composite voltage stepping through discrete levels within each PWM cycle, while the bottom STFT panel shows broadened bands and blurred boundaries: (i) short level durations are averaged with adjacent levels and may not appear as separate bands, (ii) transitions are delayed and appear spread in time rather than sharp, and (iii) closely spaced levels partially merge when their separation is comparable to the window’s effective bandwidth. As a result, STFT-based processing struggles to accurately capture frequency transitions and level durations.

\sssec{Estimating Instantaneous Frequency:} \sysname tracks the  frequency shifts by exploiting abrupt phase jumps in the backscattered signal. These jumps arise because the RF switch toggles the tag antenna's reflection coefficient between two impedances, effectively multiplying the carrier by a square wave. Each rising or falling edge of the square wave induces a sharp phase change. \sysname computes  the adjacent-sample phase difference of the received I/Q signal and finds their peaks with indices  $idx(k)$; the instantaneous frequency between successive peaks is then calculated as $f_{inst}(k) = \frac{f_s}{2(idx(k)-idx(k-1))}$, where $f_s$ is  the receiver sampling rate.  The bottom panel of Fig.~\ref{fig:inst_Freq} shows the resulting  frequency trace. 
Indeed, time-varying multipath from nearby  moving objects can perturb the phase and thus frequency estimates; we mitigate this using an HMM-based demultiplexer in Sec.~\ref{sec:HMM}.

\sssec{Trace Segmentation:} Next, \sysname must segment the frequency trace into individual PWM cycles. Each cycle begins at the lowest frequency level—the VCO maps the highest  voltage to the lowest output frequency~\cite{LTC6990}—and then steps upward as the PWM duration progress until reaching the highest frequency when all sensors reach PWM-low.
A na\"{\i}ve approach is to use a fixed threshold to mark cycle boundaries, for example, (i) declaring a new cycle when the frequency drops below a predefined level, or (ii) triggering a boundary when the high‑to‑low drop exceeds a fixed magnitude. This works on ideal clean traces but is sensitive to drift and jitter
so a single threshold either fires too often or misses the cycle start, as shown  in Fig.~\ref{fig:period}. 

To obtain reliable cycle boundaries, \sysname uses two steps: \textit{1) \underline{Coarse detection}}: identify candidate boundary indices where a pronounced high-to-low drop occurs from the highest frequency, and enforce a minimum segmentation of  $f_{min}/f_{ref}$, where $f_{min}$ and $f_{ref}$ is the lowest frequency and  RF-derived timing reference frequency, respectively. 
\textit{2) \underline{Refinement:}} 
map candidate indices  to time and compare their  spacing to the known PWM cycle duration. Boundaries that are too dense are pruned, while the gap exceeding one cycle duration is filled by inserting a boundary at the timestamp closest to the expected PWM cycle. This  aligns boundaries to the known timing reference and ensures one segmentation per PWM cycle. The red lines in Fig.~\ref{fig:period} show the corrected segmentation.

\subsection{Signal Reconstruction via HMM}\label{sec:HMM}
After segmenting the frequency trace into PWM cycles, the next task is to determine the frequency transition instants and identify the durations of each frequency level; transition instants mark changes in which sensors have PWM output high (``active''), and the durations between transitions determine per-sensor PWM duty cycles and thus sensor readings. In principle, a simple threshold-based detector could flag a transition whenever the frequency increases by more than a certain threshold (e.g., 20 kHz) as the cycle progresses from the lowest frequency level upward. However, it fails in practice due to following challenges:

\textit{1) \underline{Multipath}}: Moving objects in the surroundings introduce time-varying multipath reflections that perturb  the phase-difference-based instantaneous frequency estimates. The estimated frequency then fluctuates slightly around (see Fig.~\ref{fig:transit}(b))—rather than remain flat at  pre-allocated discrete frequency levels (set by the assigned VDM's voltage weights). 

\textit{2) \underline{Hardware Imperfections}}: Beyond the frequency jitter,  the VCO can turn a clean, sharp voltage change into a gradual glide of the backscattered frequency (shown as the red part in Fig.~\ref{fig:transit}(c)). This not only blurs the transition instant but also creates ambiguity when a portion of the duration contains both flat and transitioning segments, making it difficult to determine the exact frequency level.

Under the dual challenges of multipath and hardware imperfections, a fixed-threshold approach often produces spurious transition candidates and overestimates the number of transitions compared to the actual number of sensors (Fig.~\ref{fig:transit}(a)). A further complication arises when two or more sensors share identical readings within a cycle-their PWM high durations end simultaneously, resulting in a single transition. Thus, the actual number of correct transitions can be smaller than the number of sensors (see Case 3 and 4 in Fig.~\ref{fig:binary}). These merged transitions, coupled with spurious ones, can easily confuse a simple threshold-based detector. Therefore, \sysname must precisely identify the correct transition instants—no more than the number of sensors—from a noisy, large set of transition candidates, and achieve robust demultiplexing in the presence of spurious and merged transitions.

\sssec{Demultiplexing using HMM:} \sysname initially flags candidate transitions   with a simple threshold rule. 
To refine these candidates and jointly infer, within each PWM cycle, the sequence of frequency levels and their durations, \sysname formulates the task as a Hidden Markov Model (HMM) and solves it with  the Viterbi algorithm (Algo.~\ref{alg:viterbi}). The HMM is specifically tailored to our demultiplexing: the \textit{hidden state} represents which sensors are active (PWM-high); the \textit{transition probability} constrains how  states can evolve while accommodating hardware imperfections; and the \textit{emission probability} quantifies how well the observed frequency trace  matches each state while explicitly tolerating time‑varying multipath—modeling  deviations around the
 pre-allocated frequency levels. The Viterbi algorithm then searches for the globally most likely sequence of states 
and their durations. 
Specifically:

\textbullet\ \textit{Hidden States.} A hidden state of HMM encodes which of the $N$ sensors are active (PWM high).   We represent a state $S$ by an $N$-bit binary vector: a  \texttt{'1'} at position  $j$ means sensor $j$  is active, and \texttt{'0'} means inactive. For example,  $S=\texttt{11101}$ means sensors 4, 3, 2, and 0 are active, while sensor 1 is inactive. Let $S_i$ denote the state on the $i$-th observed frequency interval—i.e., the frequency trace between  candidate transition $i$ and  transition $i+1$.  Each cycle begins in the all-active state (all sensors are PWM-high), so the initial probability is $P_{init}(S_0 = 2^{N-1})=1$ and $0$ for other states.

\textbullet\ \textit{Transition Probability.} Within a PWM cycle, VDM's multiplexing design impose two constraints: 1) one or more sensors may drop from PWM‑high to low at the same instant, and 2) once low, a sensor does not rise again until the next period. Spurious transitions can appear due to hardware imperfections, so \sysname's state transition model uses two safeguards: the next state moves only to a subset of the current state, and when the frequency glides or the interval do not clearly belong to any subset, the state can remain unchanged, thereby allowing ambiguous transitions to be absorbed as part of current state. Specifically, let $\mathrm{Active}(x)$ denote the set of sensors that are high in state $x$. We then define  the transition probability  from  state $S_{i} = x$ to  state $S_{i+1} = y$ as:
 
\begin{equation}
P_{\text{trans}}(S_{i+1} = y \mid S_i = x) =
\begin{cases}
1, & \text{if } y = x, \\[2pt]
1, & \text{if } \mathrm{Active}(y)\subseteq \mathrm{Active}(x),\\[2pt]
0, & \text{otherwise}.
\end{cases}
\label{eq:trans_prob}
\end{equation}

\textbullet\ \textit{Emission probability.} Let $\mathrm{Seg}_i$ denote the interval of the frequency trace between candidate transitions $i$ and $i{+}1$. The emission probability $P(Seg_i \mid S_i = x)$ is the likelihood of this interval given state $x$. Within $\mathrm{Seg}_i$, we further partition the trace into \textit{slices} bounded by adjacent dots in Fig.~\ref{fig:transit};  slice $j$ has mean frequency $f_j$. Due to varying multipath, $f_j$ may fluctuate and deviate from the pre-allocated discrete frequency level. Our key observation is that, for a given state $x$, slice frequencies  cluster around the pre-allocated frequency $\text{Freq}(x)$. Therefore, we model the distribution of the slice frequency $f_j$ for state $x$ as a Gaussian centered at $\text{Freq}(x)$: $P(f_j \mid S_i = x) = \mathcal{N}(f_j; \mu=\text{Freq}(x), \sigma)$, where $\sigma$ is set to the minimum inter-level frequency spacing. To obtain the emission probability of an interval that is robust to hardware-induced frequency  glides, we  aggregate probability over \textit{slices} using a weighted median across time:
$$
P(Seg_i \mid S_i = x) = \mathrm{Median}\left( W_{\text{slice}[j]} \cdot P(f_j \mid S_i = x) \right),
$$ 
where $W_{\text{slice}[j]}$ is the duration of slice $j$. The median is used in place of the mean to improve robustness to outliers.

\sssec{Reconstruction Summary:} With the HMM, \sysname applies a Viterbi-based dynamic programming (detailed in Algorithm.~\ref{alg:viterbi}) to infer the most probable transition sequence from the observed intervals based on maximizing the joint probability. Candidate intervals with spurious transitions receive lower emission probabilities and are absorbed  as self-transitions, effectively filters out spurious state changes while preserving valid ones. The final output is the sequence of transition states and the duty cycle of each sensor for a particular PWM cycle. We then map the duty cycle back to the per-sensor readings by multiplying the maximum amplitude $V_{max}$ of the known RF-derived sawtooth reference.

\begin{algorithm}[t]
	\setstretch{0.5}
	\caption{Demultiplexing via HMM}
	\label{alg:viterbi}
	\KwIn{\texttt{Intervals}: $[Seg_0, Seg_1, \dots, Seg_{M-1}]$ extracted from transition points}
	\KwOut{\texttt{Path}: Optimal state sequence, \texttt{Ratio}: duty cycle of each state}
	
	\texttt{States} $\leftarrow$ $[0, 2^N)$
	
	\texttt{M} $\leftarrow$ \texttt{len(Segments)}
	
	\textbf{\# Initialization}
	
	\texttt{V[0][$2^N-1$]} $\leftarrow$ $1$ \tcp{\texttt{V[i][j]} is the max probability of path ending in state \texttt{j} at interval \texttt{i}}
	
	\textbf{\# DP Forward Pass}
	
	\For{\texttt{i} $\leftarrow$ 1 \KwTo \texttt{M-1}}{
		\ForEach{\texttt{y} in \texttt{States}}{
			\texttt{x*} $\leftarrow$ $\arg\max_{x} \left( \texttt{V[i-1][x]} \cdot P(S_i = y \mid S_{i-1} = x) \right)$ \;
			\texttt{V[i][y]} $\leftarrow$ $\texttt{V[i-1][x*]} \cdot P(S_i = y \mid S_{i-1} = \texttt{x*}) \cdot P(Seg_i \mid S_i = y)$ \;
			\texttt{prev[i][y]} $\leftarrow$ \texttt{x*} \;
		}
	}
	
	\textbf{\# Backtracking}
	
	\texttt{Path} $\leftarrow$ $[\,]$ \;
	\texttt{Ratio} $\leftarrow$ $[\,]$ \;
	\texttt{s} $\leftarrow$ $\arg\max_{x} \texttt{V[M-1][x]}$ \;
	\For{\texttt{i} $\leftarrow$ \texttt{M-1} \KwTo \texttt{0}}{
		\texttt{Path.insert(0, s)} \;
		\texttt{Ratio.insert(0, len(Seg\_i))} \tcp{Ratio based on segment duration}
		\texttt{s} $\leftarrow$ \texttt{prev[i][s]} \;
	}
	
	\textbf{\# Merge self-loop states:}
	
	\texttt{Path, Ratio} $\leftarrow$ \texttt{merge(Path, Ratio)}
	
	\Return{\texttt{Path, Ratio}}
	
\end{algorithm}


\section{Implementation}



\begin{figure*}[t]
	\centering
	\subfigure[\sysname tag]{
		\centering
		\includegraphics[width=0.28\linewidth]{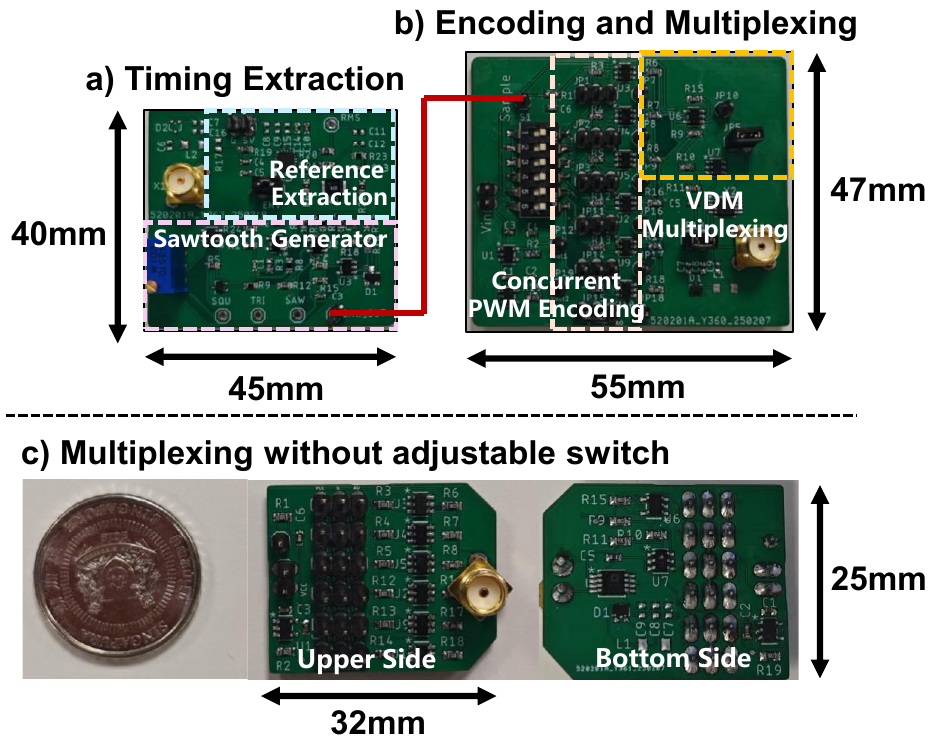}
		\label{fig:Tag}
	}\hspace{10pt}
	\subfigure[\sysname Reader]{
		\centering
		\includegraphics[width=0.28\linewidth]{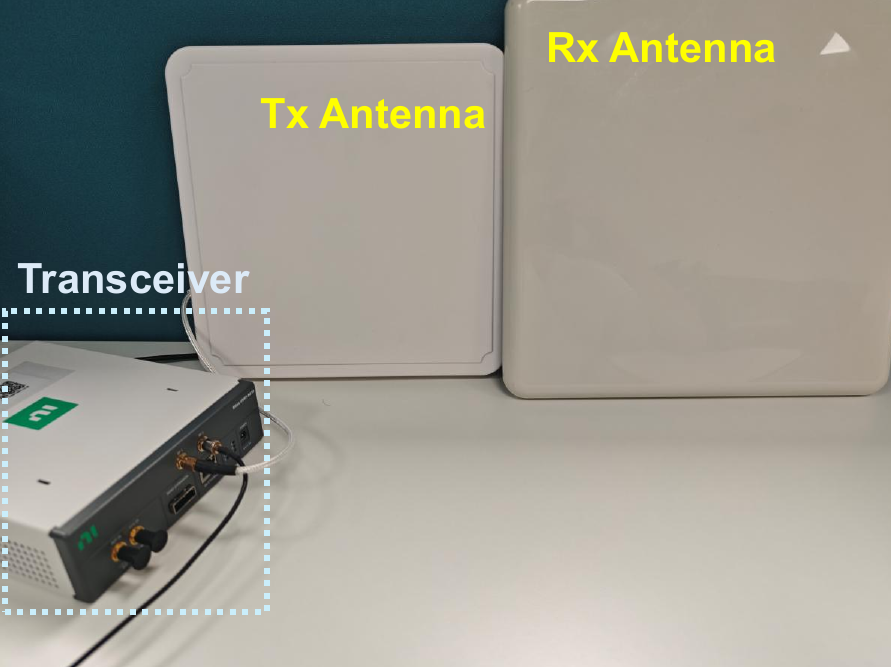}
		\label{fig:expe_setup}
	}\hspace{10pt}
	\subfigure[Floor plan]{
		\centering
		\includegraphics[width=0.32\linewidth]{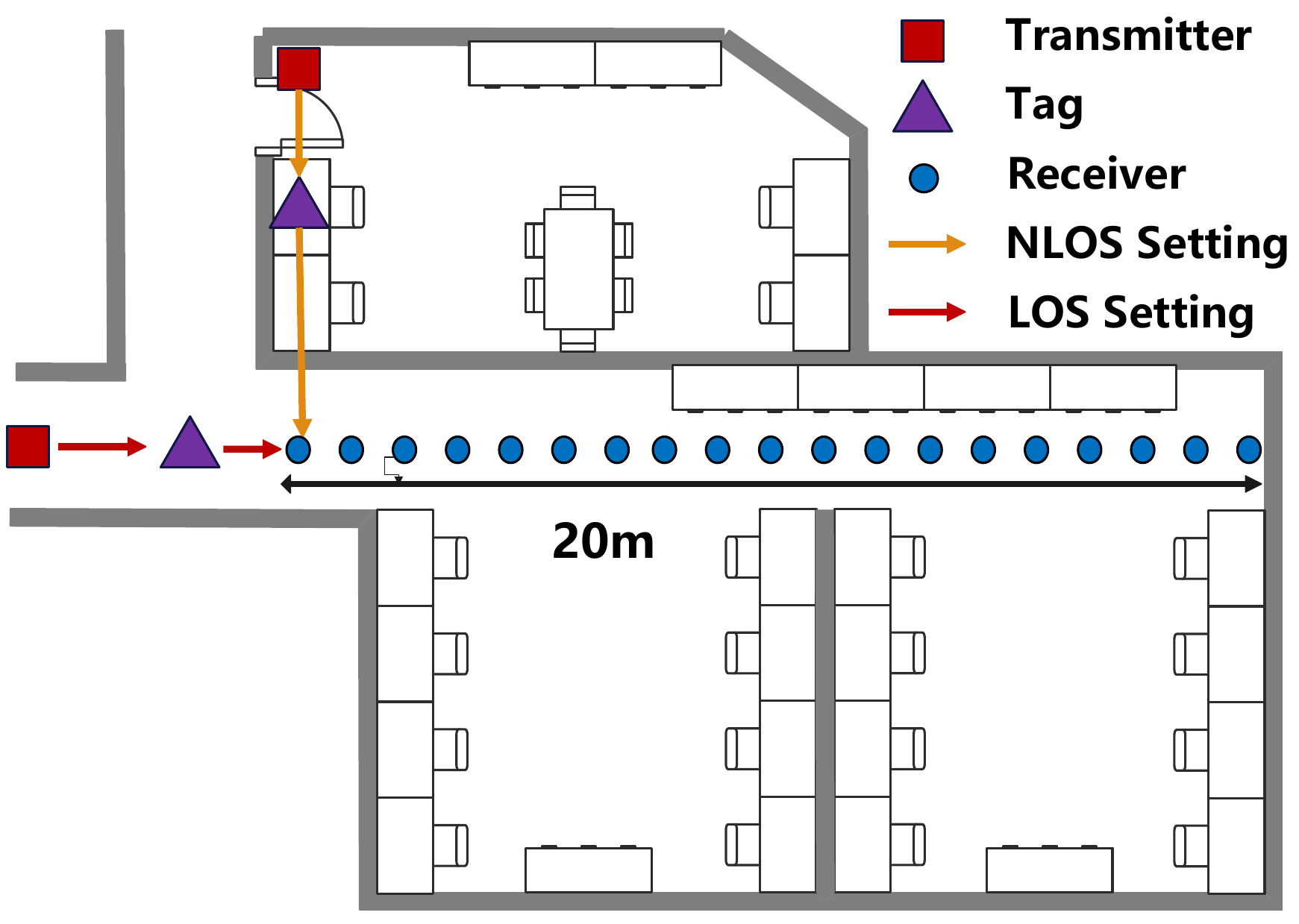}
		\label{fig:floorplan}
	}
	\vspace{-15pt}
	\caption{(a) A \sysname tag consists of two  components: 1) Sawtooth reference generation, which extracts timing signals from ambient RF excitation, and 2) Sensor encoding, multiplexing and backscatter transmission. These two components are wired together using a jumper wire. (b) \sysname reader includes one TX antenna and one RX antenna. (c) Floorplan. } 
	\label{fig:expe}
	\vspace{-10pt}
\end{figure*}

\sssec{\sysname Tag:} We implement  \sysname prototype using   off-the-shelf  analog hardware components to ensure low-power consumptions. For evaluation purposes, we add a five-way switch onboard to select how many sensors are enabled concurrently; without the switch, the tag fits in $32mm \times 25mm$ as shown in Fig.~\ref{fig:Tag}.   \sysname uses two  low-cost 915 MHz antennas, each with a 3 dBi gain: one  for  backscatter and one to derive sawtooth timing reference, orthogonally polarized. The cost of a  \sysname tag is around USD 17. The two-layer PCB prototype draws $1.36 mW \sim 2.32 mW$, depending on onboard sensor count (1–5) and the sampling frequency.

\sssec{\sysname Reader:} We use a USRP N210 SDR equipped with an SBX-40 daughterboard~\cite{SBX-40} as a 915MHz-band RF transceiver to send a two-tone carrier signal and receive backscattered signal from the tag. Two 9 dBi polarized antennas are used for transmission (TX) and reception (RX). The transmission power is 20 dBm and the USRP's digitization (sampling) rate is set to 20 $MHz$.

\sssec{\sysname Software:} All processing runs at the \sysname reader. The RF signal is continuously demultiplexed into individual sensor readings on a 2025 MacBook Air M4, 16GB RAM, with the demultiplexing algorithm entirely implemented in Python 3.9. For 5 concurrent sensors at a sampling frequency (tag's time base) of 50kHz, the computational latency for demultiplexing   is $0.97 ms$ per sampling period (one PWM cycle).

\sssec{IC Design:} We design ASIC of \sysname using Cadence IC6.17 Virtuso. The design is implemented  in the Virtuso analog design environment using the TSMC 65nm CMOS Low Power technology library. The ASIC design shows that the overall power consumption for timing reference extraction and signal multiplexing is 7.47 $\mu W$ and 18.09 $\mu W$, respectively. The power consumption of each component's prototype is described in Sec.~\ref{sec:power}.    

\section{Evaluation}\label{sec:eval}

\subsection{Evaluation Methodology}


\sssec{Ground Truth:} For \sysname's overall-capability evaluation (e.g., supported sampling frequency, sensor signal frequency) in Sec.~\ref{sec:eval}, we emulate sensor outputs with a synchronized five-channel signal generator assembled from three Analog Discovery 3 units~\cite{AD3}. It produces programmable voltages from (0-3.3 $V$)  and signals  up to 24 kHz, covering the typical voltage levels and bandwidths of analog sensors. These signals serve as the ground-truth input to \sysname.
Note that for the case studies in Sec.~\ref{sec:application}, we interface real sensors to the tag and collect live signals as the reference.

\begin{figure*}
	\centering
	\begin{minipage}{0.23\linewidth}
		\includegraphics[width=\linewidth]{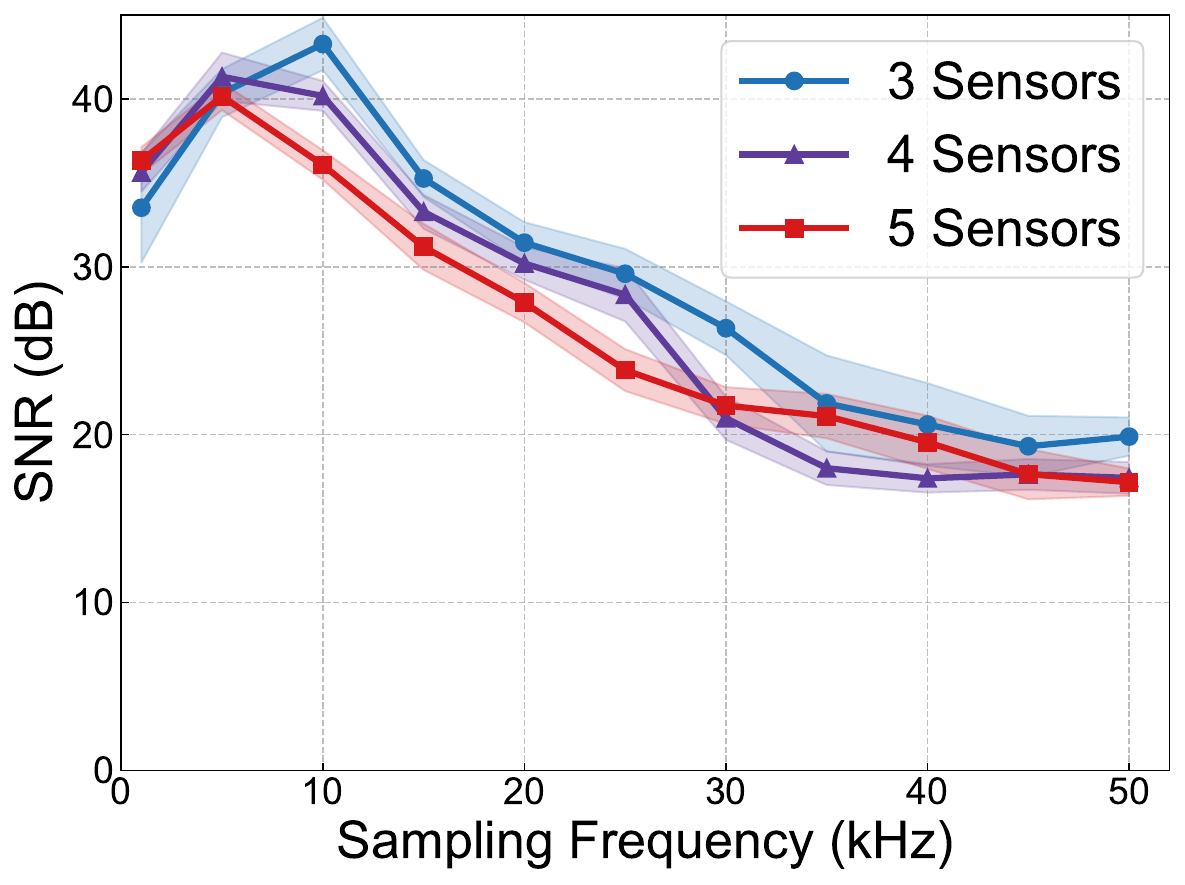}
		\vspace{-20pt}
		\caption{SNR vs. Sampling frequency.} 
		\label{fig:ref_freq}
	\end{minipage}\hspace{10pt}
	\begin{minipage}{0.23\linewidth}
		\includegraphics[width=\linewidth]{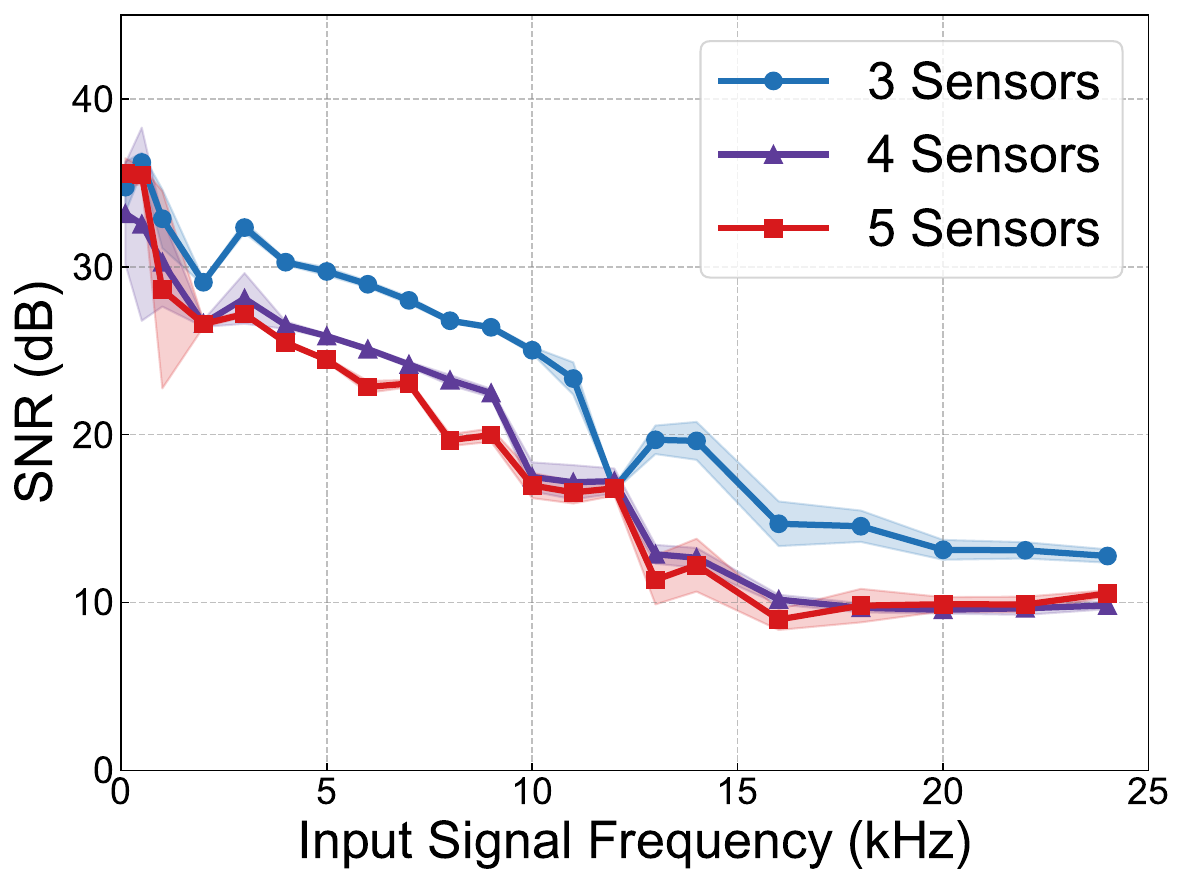}
		\vspace{-20pt}
		\caption{SNR vs. sensor input frequency.} 
		\label{fig:sample_Freq}
	\end{minipage}\hspace{10pt}
	\begin{minipage}{0.23\linewidth}
		\includegraphics[width=\linewidth]{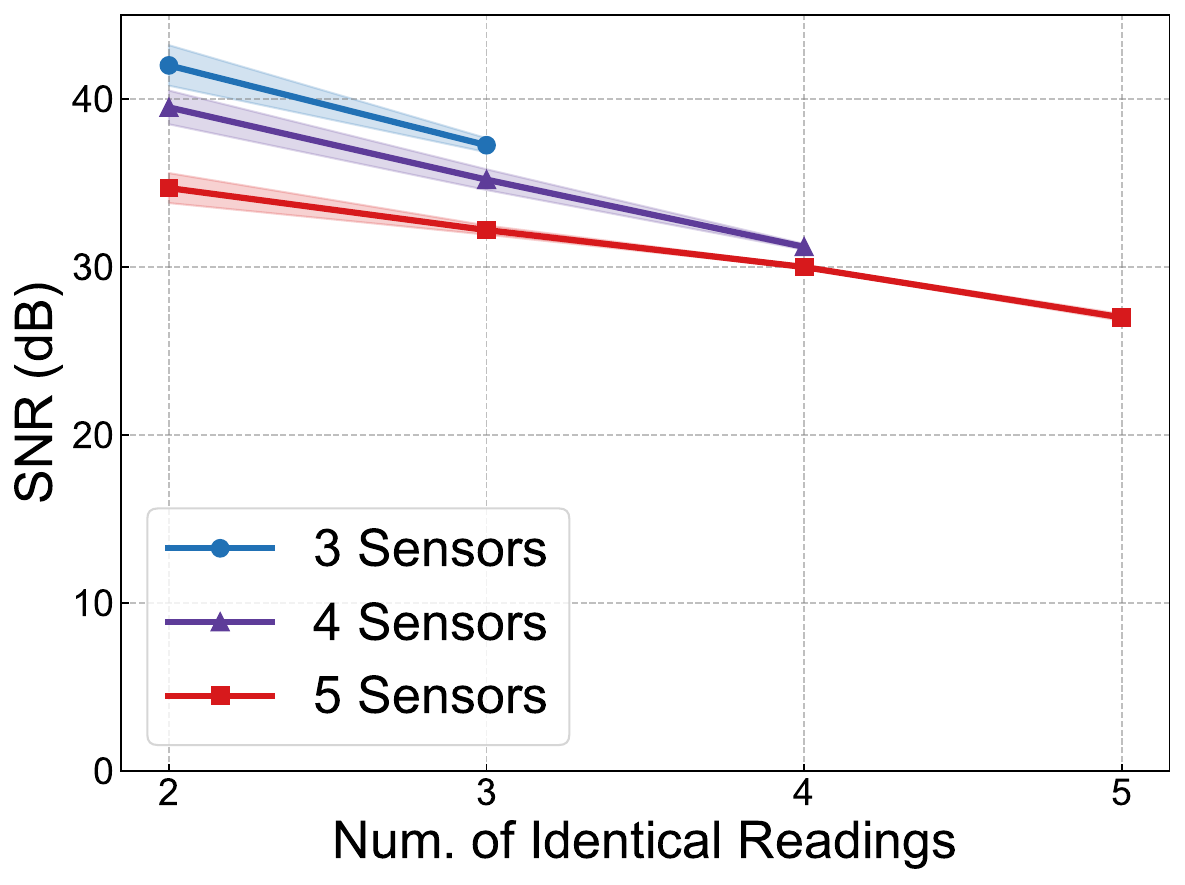}
		\vspace{-20pt}
		\caption{SNR vs. the number of identical readings.} 
		\label{fig:identical}
	\end{minipage}\hspace{10pt}
	\begin{minipage}{0.23\linewidth}
		\includegraphics[width=\linewidth]{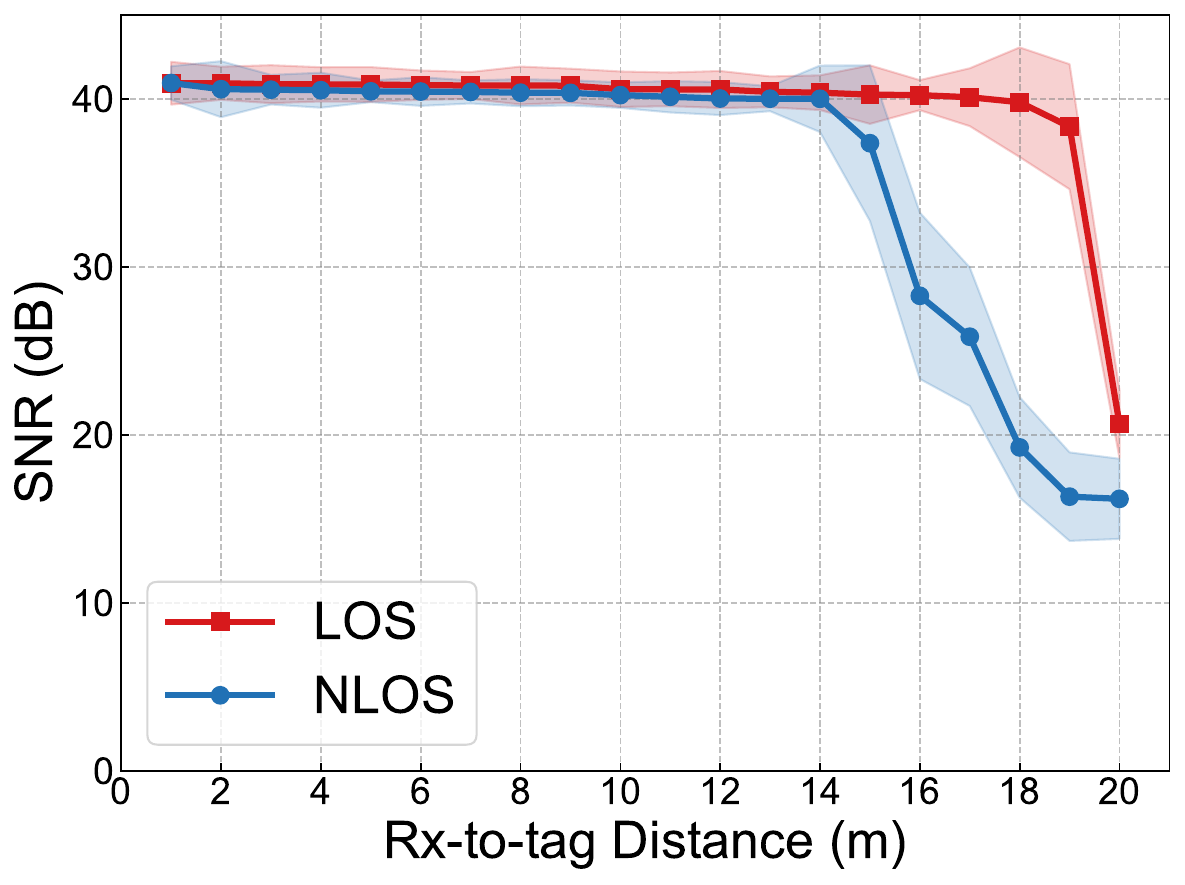}
		\vspace{-20pt}
		\caption{SNR under impact of Rx-to-Tag distance.} 
		\label{fig:rxtotag}
	\end{minipage}\hspace{10pt}
\end{figure*}

\begin{figure*}
	\centering
	\vspace{-10pt}
	\begin{minipage}{0.23\linewidth}
		\includegraphics[width=\linewidth]{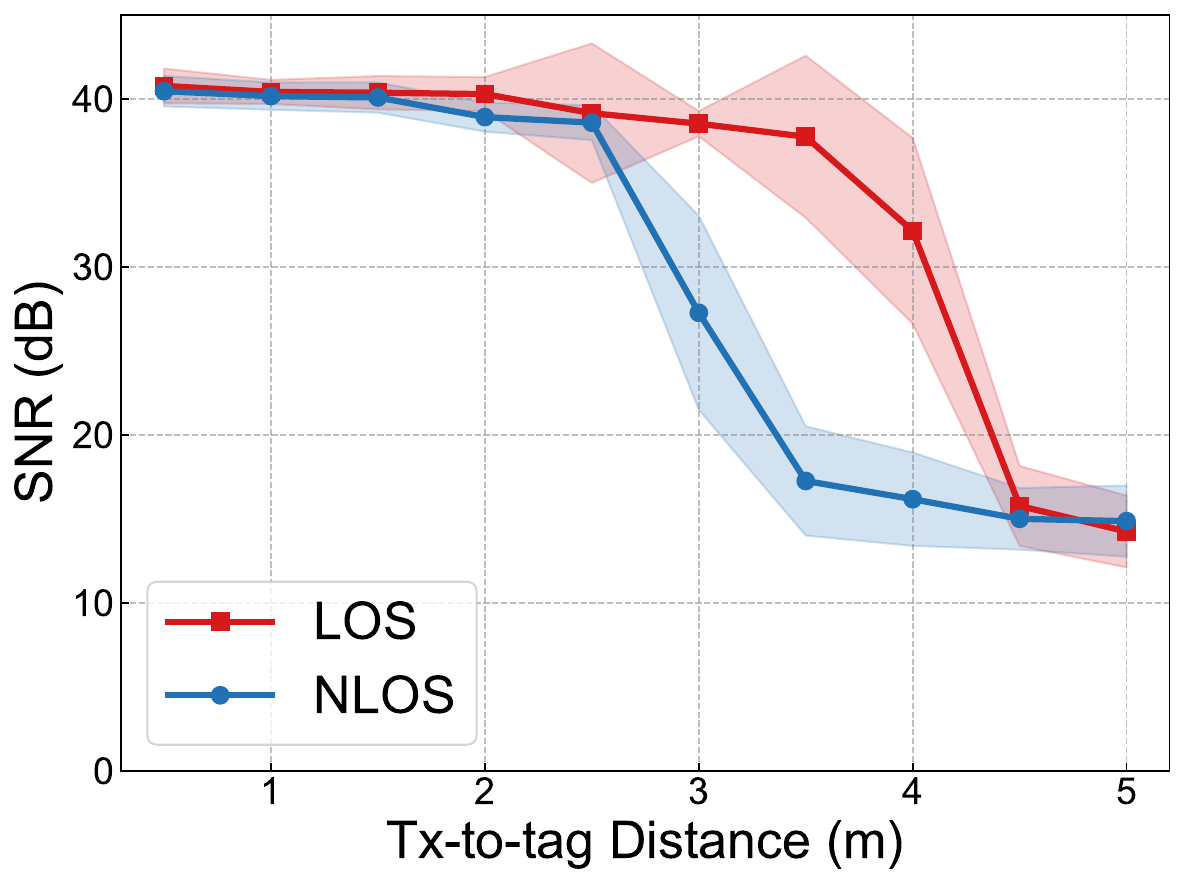}
		\vspace{-20pt}
		\caption{Impact of Tx-to-Tag distance.} 
		\label{fig:txtotag}
	\end{minipage}\hspace{10pt}
	\begin{minipage}{0.23\linewidth}
		\centering
		\includegraphics[width=1\linewidth]{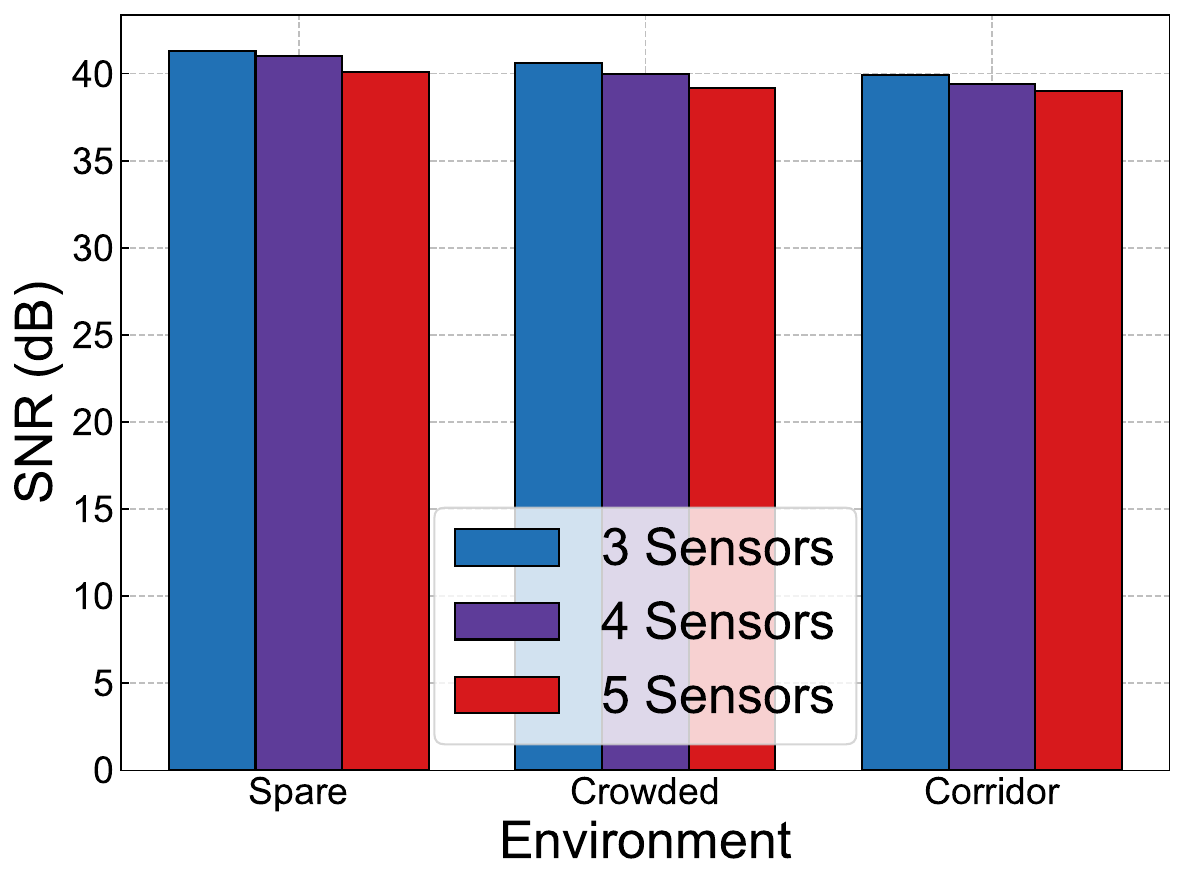}
		\vspace{-20pt}
		\caption{Various Environmental Dynamics.} 
		\label{fig:envi_int}
	\end{minipage}\hspace{10pt}
	\begin{minipage}{0.23\linewidth}
		\includegraphics[width=\linewidth]{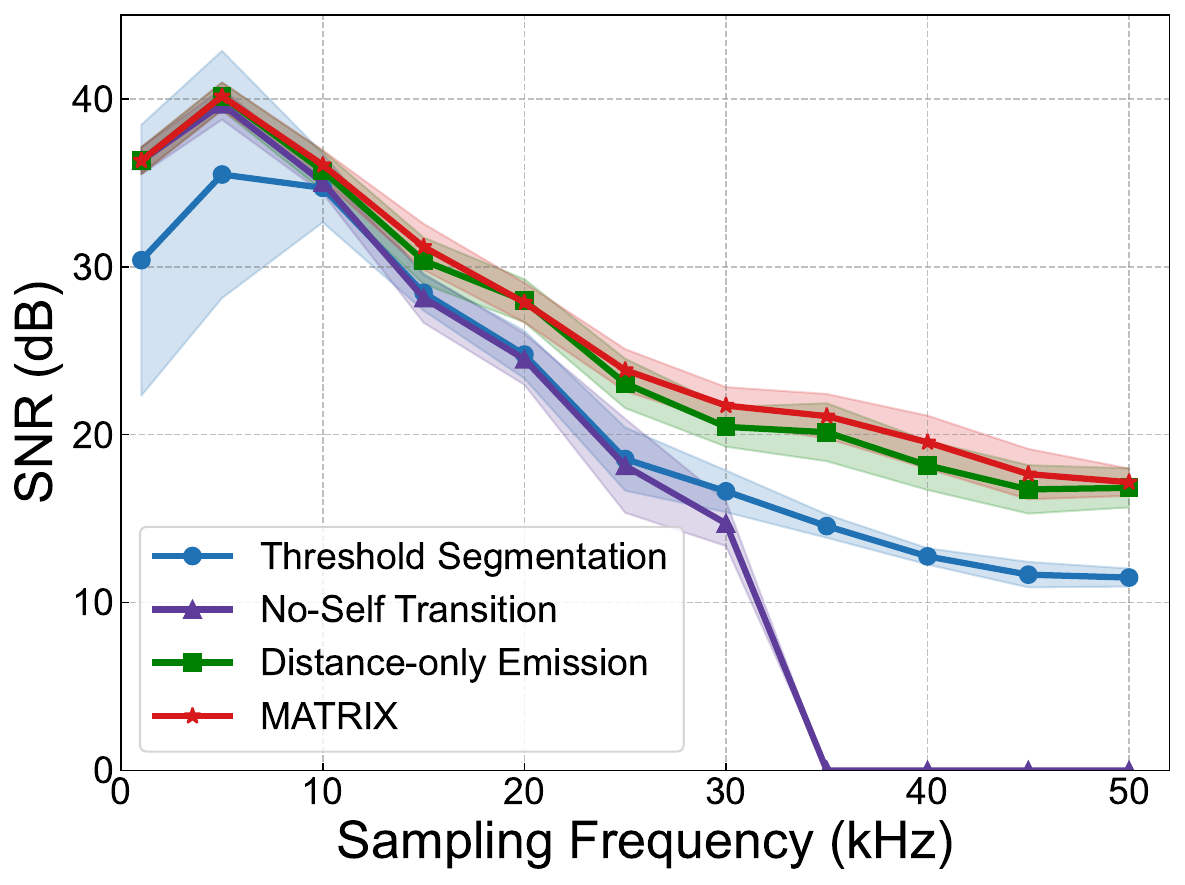}
		\vspace{-20pt}
		\caption{Ablation across sampling frequencies.} 
		\label{fig:algo_samp}
	\end{minipage}\hspace{10pt}
	\begin{minipage}{0.23\linewidth}
		\includegraphics[width=\linewidth]{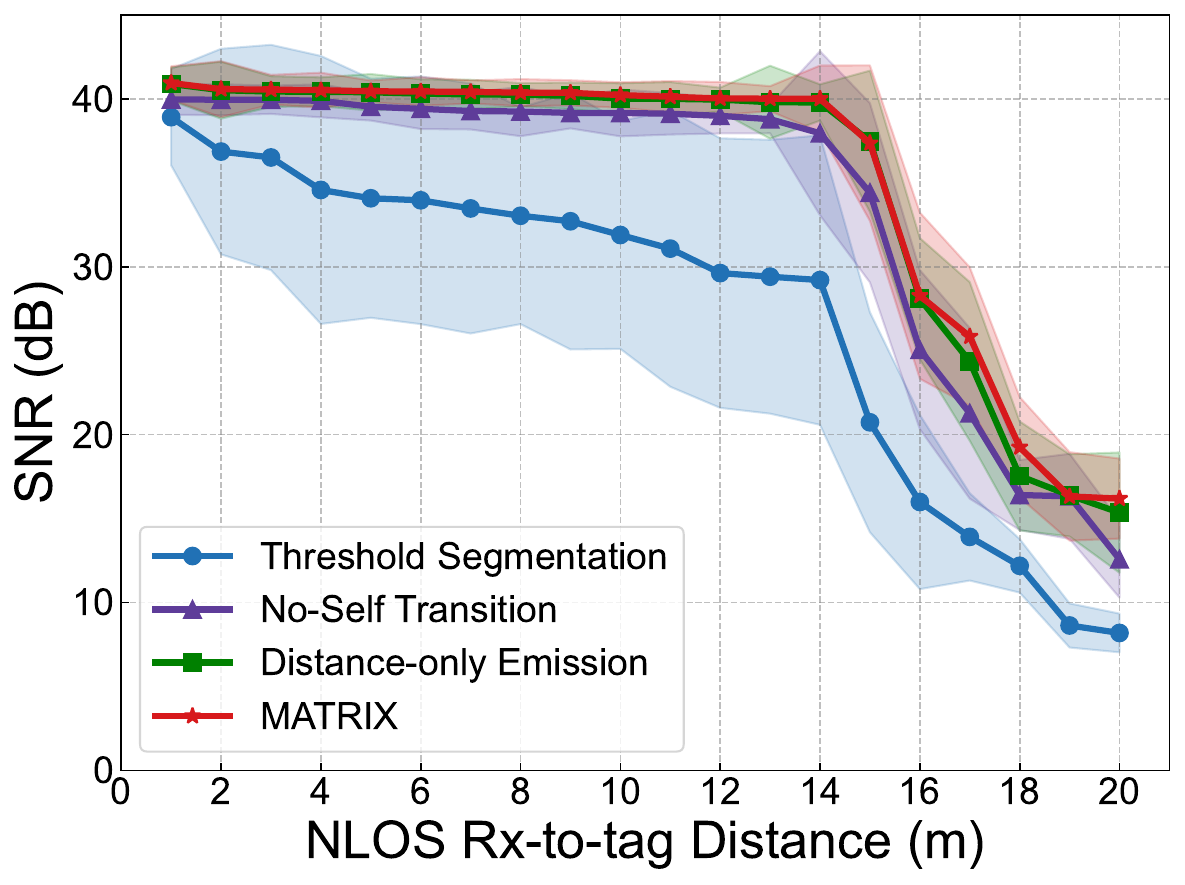}
		\vspace{-20pt}
		\caption{Ablation across Rx-to-Tag distance.} 
		\label{fig:algo_dist}
	\end{minipage}\hspace{10pt}
	\vspace{-10pt}
\end{figure*}



\sssec{Environments:} The experiments were conducted indoors across multiple rooms furnished with tables and chairs (floor plan in Fig.~\ref{fig:floorplan}). We evaluated both line-of-sight (LOS) and non-line-of-sight (NLOS) conditions, all with moving objects present.

\sssec{Evaluation Metric:} We quantify signal reconstruction quality using Signal-to-Noise Ratio (SNR in dB), defined as $SNR(x,\hat{x})=10\log_{10}(\frac{||x||_2^2}{||x-\hat{x}||_2^2})$, where $x \in R^{1\times T}$ and $\hat{x} \in R^{1\times T}$ denote the ground-truth signal and reconstructed signals. We reported the average SNR across 30 sequences and all onboard sensors. Note that SNR is used instead of bit error rate (BER) since \sysname carries continuous analog sensor waveforms rather than framed discrete bitstream.

\subsection{SNR vs. Sampling Frequency}\label{subsec:samplingfreq}

The sampling frequency is the shared timing reference (derived from the ambient carrier) used to convert each sensor value into a PWM waveform; it sets the PWM-cycle duration–think of it as the sampling rate of the raw sensor signals. 
With a fixed RF digitization rate at the receiver (e.g., 20 MHz at USRP), increasing the sampling frequency shortens each PWM cycle and thus reduces the number of digitized RF samples  per cycle.  This indeed poses challenges for  demultiplexing, as fewer digitized RF samples per cycle make it harder to pinpoint frequency transitions and their durations, raising the difficulty of signal reconstruction. We therefore evaluate \sysname's demultiplexing performance across a wide range of sampling frequency.

\sssec{Method:} We evaluate \sysname's  SNR  across sampling frequency from 1 kHz to 50 kHz for configurations  with 3–5 concurrent sensor streams.  Individual sensor signals are emulated with the signal generator; each is  a 100 $Hz$ single-tone sinusoid with random phase and amplitude. The Tx-to-tag  and the Rx-to-tag distances are $1m$ and $5m$; the impact of varying distances is evaluated in Sec.~\ref{sec:distance}. 

\sssec{Results:} Fig.~\ref{fig:ref_freq} shows the average SNR vs. sampling frequency for 3–5 concurrent sensor streams. At low sampling frequencies (below 10 $kHz$), 
SNR rises modestly with sampling frequency because the accurate signal reconstruction benefits from finer timing resolution in the raw sensor signals while still receiving ample digitized RF samples per PWM cycle; 
peaks occur near 10 $kHz$ for 3 sensors and 5 $kHz$ for 4–5 sensors, typically above 30 $dB$. As sampling frequency increases further, each cycle contains fewer digitized RF  samples, raising the difficulty of signal reconstruction and causing SNR to decline. At a target SNR of 20 $dB$, the 3-sensor setting sustains sampling up to 50 $kHz$, whereas 4/5-sensor settings sustain up to 30 $kHz$. We note that at 50 kHz, the 4/5-sensor settings still reach 18 $dB$ SNR. The 5-sensor setting occasionally outperforms 4-sensor one but shows significantly larger variance. 

\subsection{SNR vs. Sensor Signal Frequency}\label{sec:sensorfreq}

\sssec{Methods:} Real-world sensors produce outputs across diverse frequency range. We therefore evaluate the impact of the sensor signal frequency on SNR.  We generate sensor frequency  from 100 $Hz$ to 24 $kHz$ and measure the average SNR for 3–5 concurrent sensors. We set the sampling frequency slightly above twice the sensor signal frequency (just above Nyquist rate) so that each  signal cycle spans about 2–3 PWM cycles,  ensuring fair and consistent sampling conditions across  sensor frequencies.  Distances are as in  Sec.~\ref{subsec:samplingfreq}.

\sssec{Results:} Fig.~\ref{fig:sample_Freq} shows that SNR remains above 20 $dB$ up to 10 $kHz$ input frequency for all 3–5-sensor settings, demonstrating robust performance.
We observe that SNR decreases as the sensor signal frequency increases. At around 24 $kHz$, \sysname achieves about 12 $dB$ for 3-sensor, and 9-11 $dB$ for 4/5-sensor settings.

\subsection{SNR vs. Number of Identical Readings}\label{sec:identical}

\sssec{Methods:} Identical readings across onboard sensors can merge PWM-high transitions (see  Fig.~\ref{fig:binary}). We evaluate the \sysname's performance in handling these merges. We generate the sensor signals at $100Hz$ and sample at $10kHz$, and test 3-5 concurrent sensor streams. We start with two identical readings, then gradually align each sensor's amplitude and phase until all match  the first sensor. The TX-to-tag and the RX-to-tag distances are $1m$ and $5m$, respectively. 

\sssec{Results:} Fig.~\ref{fig:identical} shows that as more readings become identical,  SNR gradually decreases. This degradation occurs because merged transitions create longer frequency glides between frequency levels,  resulting in less precision in pinpointing transitions and introducing  distortions in the amplitude of the reconstructed signals. Nevertheless,  SNR remains above 28 $dB$ in all cases, demonstrating the robustness to the merged transitions.

\subsection{SNR vs. Distances}\label{sec:distance}
This section evaluates the impact of the relative placement of the transmitter antenna (Tx), receiver antenna (Rx) and the tag under both LOS and NLOS conditions. The sampling frequency is 5 $kHz$ and the sensor signal frequency is 100 $Hz$. 
 
\sssec{Rx-to-Tag Distance:}\label{subsec:rxtotag}
We fix the Tx-to-Tag distance at 1 $m$. \sysname tag operates 5 onboard sensors. The Rx is placed 1–20 $m$ from the tag in 1 $m$ increments. For each Rx-to-Tag setup under both LOS and NLOS conditions (see Fig.~\ref{fig:floorplan}), we collect 5 runs and measure the SNR.  Fig.~\ref{fig:rxtotag} shows that SNR remains around 40 $dB$ up to 14  $m$ in both LOS and NLOS conditions.  This stability arises because \sysname's instantaneous frequency tracking, based on phase differences,  is robust to  distance-induced amplitude variations,  as long as the received signal stays above the noise floor. 
Once the received signal strength falls below the noise floor at longer distances (19 $m$ for LoS, and 15 $m$ for NLoS), performance drops sharply.  Overall,  with five sensors, \sysname maintains at least 20 $dB$ SNR up to 18 $m$ in both LOS and NLOS.



\sssec{Tx-to-Tag Distance:} We fix the Rx-to-Tag distance at 5 $m$, operate five sensors, and vary Tx-to-Tag distance from 0.5 $m$ to 5 $m$ in $0.5m$ increments under both LOS and NLOS scenarios.  Fig.~\ref{fig:txtotag} shows that SNR decreases with increasing  Tx-to-Tag distance, from 40 $dB$ at  0.5 $m$ to 15 $dB$ at 5 $m$. The  LOS curve drops more gradually and remains consistently above the NLOS. Overall, with a 20 $dBm$ Tx power, SNR  stays $>$ 20 dB out to 4 $m$ in LoS and 3 $m$ in NLOS. 


\subsection{Robustness to Environmental Dynamics} 

\sssec{Method:} Movement  in the surroundings introduces time-varying multipath. We evaluate \sysname’s robustness to such environmental dynamics. We deploy \sysname in three scenarios: 1) a spacious, sparsely furnished  room, 2) a crowded room furnished with tables and chairs, and 3) a busy corridor with pedestrian traffic. In this experiment, the Tx-to-Tag distance is $1m$ and the Rx-to-Tag distance is $10m$, allowing people to walk through the signal propagation path. The sampling frequency is 5 $kHz$ and the sensor signal frequency is 100 $Hz$.   For each scenario, we test 3–5 concurrent sensor streams.

\noindent\textbf{Results:} Fig.~\ref{fig:envi_int}  shows that \sysname  remains robust   across environments. The  spacious  environment provides  about 1 $dB$ higher SNR  than multipath-rich environments, but all scenarios  exceed 35 $dB$ regardless of the number of sensors. These results indicate that \sysname's demultiplexer  is resilient to time-varying multipath caused by movement in the  environment. 

	
\subsection{Ablation Study} 

\sysname uses a two-stage trace segmentation to reliably locate PWM-cycle boundaries and an HMM demultiplexer to infer, within each cycle, the sequence of frequency levels and their durations—both designed to be robust to analog hardware imperfections and multipath. We quantify the effectiveness of these design choices by comparing \sysname against baselines  discussed in Sec.~\ref{sec:demodulation}.

\sssec{Method:} We run two testing setups in multipath-rich conditions:  (i) sampling frequency varied from 1 to 50 kHz (setup per Sec.~\ref{subsec:samplingfreq}) and (ii) Rx-to-Tag distance varied from 1 to 20 m under NLoS (setup per Sec.~\ref{subsec:rxtotag}).  We operate five sensors and report average SNR. We compare against three baselines: (a) \textit{Threshold Segmentation:} replaces our two-stage segmentation with a fixed‑threshold PWM-cycle detector; (b) \textit{No Self-Transition:} disables state self‑transition in the HMM; (c) \textit{Distance-only Emission:} replaces the Gaussion emission with  the distance to the pre-defined frequency levels.


\sssec{Results:} \textbf{(1) Ablation across sampling frequency.} Fig.~\ref{fig:algo_samp} shows that \sysname achieves  the highest SNR across all sampling frequencies. We note that the \textit{distance‑only emission} baseline  is close to \sysname at low sampling frequency, but degrades at high sampling frequency. This is  because, as the sampling frequency increases, the VCO and RF switch must sweep across the spectrum faster to keep up with the shorter PWM cycles, which increases susceptibility to  multipath-induced frequency jitter. \sysname's Gaussian emission is specifically designed to tolerate these frequency deviations. We further observe that the \textit{no self-transition} baseline  drops sharply beyond 30 $kHz$  because more salient hardware-induced frequency transition glides trigger spurious state changes that then propagate errors to the next states, whereas \sysname's self-transition absorbs ambiguous transitions (see Sec.~\ref{sec:HMM}).
Finally,  the \textit{threshold segmentation} consistently performs worse than \sysname's two-stage segmentation   as it tends to over‑/under‑cut of PWM cycles.

\textbf{(2) Ablation across Rx-to-Tag distance.} \sysname outperforms 
the  baselines and remains near 40 dB SNR up to 14 $m$,  while the \textit{threshold segmentation} degrades increasingly with   distance.
We note that at a 5 kHz sampling frequency, the \textit{threshold segmentation} is more susceptible to frequency fluctuations induced by multipath, resulting in dense cuts within a PWM cycle;  this  also yields substantially higher variance than \sysname and  other baselines.

%
%
%


\begin{table}[t]
	\renewcommand{\arraystretch}{0.8}
	\centering
	\caption{Power consumption of \sysname in both PCB and ASIC design.}
	\vspace{-10pt}
	\scalebox{0.9}{\begin{tabular}{ccc|ccc}
		\toprule[1.5pt]
		\multicolumn{3}{c|}{\textbf{Sawtooth Generator}} &
		\multicolumn{3}{c}{\textbf{Multiplexing}} \\
		\midrule[1pt]
		Freq & PCB & IC &  Num. & PCB & IC  \\
		\midrule[1pt]
		5kHz & 191.26$\mu W$ & 7.375 $\mu W$ & 1 &  1.14 $mW$ & 10.68 $\mu W$\\
		10kHz & 230.26$\mu W$ & 7.423 $\mu W$ & 2 &  1.29 $mW$ & 12.77 $\mu W$ \\
		20kHz & 312.49$\mu W$ & 7.454 $\mu W$ & 3 &  1.45 $mW$ & 14.76 $\mu W$ \\
		30kHz & 394.35$\mu W$ & 7.463 $\mu W$ & 4 &  1.60 $mW$ & 16.59 $\mu W$ \\
		40kHz & 466.81$\mu W$ & 7.467 $\mu W$ & 5 &  1.78 $mW$ & 18.09 $\mu W$ \\
		50kHz & 542.51$\mu W$ & 7.47 $\mu W$ &  \\
		\bottomrule[1.5pt]
	\end{tabular}}
	\label{tab:power}
	\vspace{-10pt}
\end{table}

\subsection{Power Consumption}\label{sec:power}
We evaluate the power consumption of \sysname tag under a prototype designed for 3.3 $V$ supply, so it can plug‑and‑play with off‑the‑shelf sensor modules (e.g., Adafruit, SparkFun). All numbers exclude the power drawn by the sensors themselves and refer only to the tag electronics.  The tag has two  blocks: i) a sawtooth time-reference extractor, and ii) the multiplexing and backscatter chain. Table~\ref{tab:power} reports the breakdown  vs. sampling frequency and the number of onboard sensors. 
	
\sssec{PCB Prototype (3.3 $V$):}The sawtooth block consumes $191$–$543~\mu W$ (from $5$–$50~kHz$ sampling), while the multiplexing block draws $1.14$–$1.78~mW$ for $1$–$5$ sensor streams.; per-branch comparators consume $127 \mu W$ while the operational amplifier dominates with $600 \mu W$. The VCO  consumes $320 \mu W$ for backscatter. 

\sssec{Path to Lower Power:} The dominant contributor to  PCB  power is operating at  3.3 V  to accommodate  the  plug-and-play sensors  used in our experiments. With  integrated low-power MEMS or bare-die sensors,  the tag can operate at a lower power (e.g., $1.2V$), significantly reducing  overall power consumption. Furthermore, by leveraging modest energy-harvesting sources (e.g., solar~\cite{parks2014turbo} or acoustic~\cite{Afzal2022, naeem2024seascan}), along with intermittent operation via a supercapacitor~\cite{sample2008design}, fully battery-free operation of \sysname is feasible. 

\sssec{ASIC: }We optimize the power consumption in the ASIC simulation. The sawtooth block consumes $7.47 \mu W$ at a 50 kHz sampling frequency. The multiplexing  consumes $10.68 \mu W$ to $18.09 \mu W$  for $1$ to $5$ sensors. In total, ASIC  consumes  $25.56 \mu W$ for 5 sensor streams at a $50 kHz$ sampling frequency.



\section{Application Use Cases}\label{sec:application}
In this section, we showcase three representative real-world applications of \sysname covering $10$ different types of off-the-shelf  sensors. The sensors tested are listed in Table.~\ref{tab:sensor_spec}.

\begin{table}[t]
	\renewcommand{\arraystretch}{0.7}
	\centering
	\caption{Sensor specifications in applications. }
	\vspace{-10pt}
	\scalebox{0.75}{\begin{tabular}{lll}
			\toprule[1.5pt]
			\textbf{Sensor} & \textbf{Part Number} & \textbf{Manufacturer} \\
			\midrule
			Water Level & 4965 & Adafruit Industries LLC ~\cite{Adafruit4965} \\
			Soil Moisture & SEN0114 & DFRobot ~\cite{DFRobotMoistureSensor} \\
			Temperature & DS18B20 &  Arduino ~\cite{ArduinoSensorKit}\\
			Light & GL5516 &  Arduino ~\cite{ArduinoSensorKit}\\
			Microphone & KY037 &  Arduino ~\cite{ArduinoSensorKit}\\
			PPG & SEN0203 & DFRobot ~\cite{DFRobotSEN0203} \\
			ECG & SEN0213 &  DFRobot ~\cite{DFRobotSEN0213}\\
			ABP &  MPS20N0040D-S & Reland Sun ~\cite{MPS20N0040DSDatasheet} \\
			RESP & SFM3020 &  Sensirion AG ~\cite{SensirionSFM3020}\\
			3-Axis Accelerometer & ADXL335 &  Analog Devices ~\cite{ADXL335}\\
			\bottomrule[1.5pt]
	\end{tabular}}
	\label{tab:sensor_spec}
	\vspace{-10pt}
\end{table}

\begin{figure}[t]
	\centering
	\includegraphics[width=0.95\linewidth]{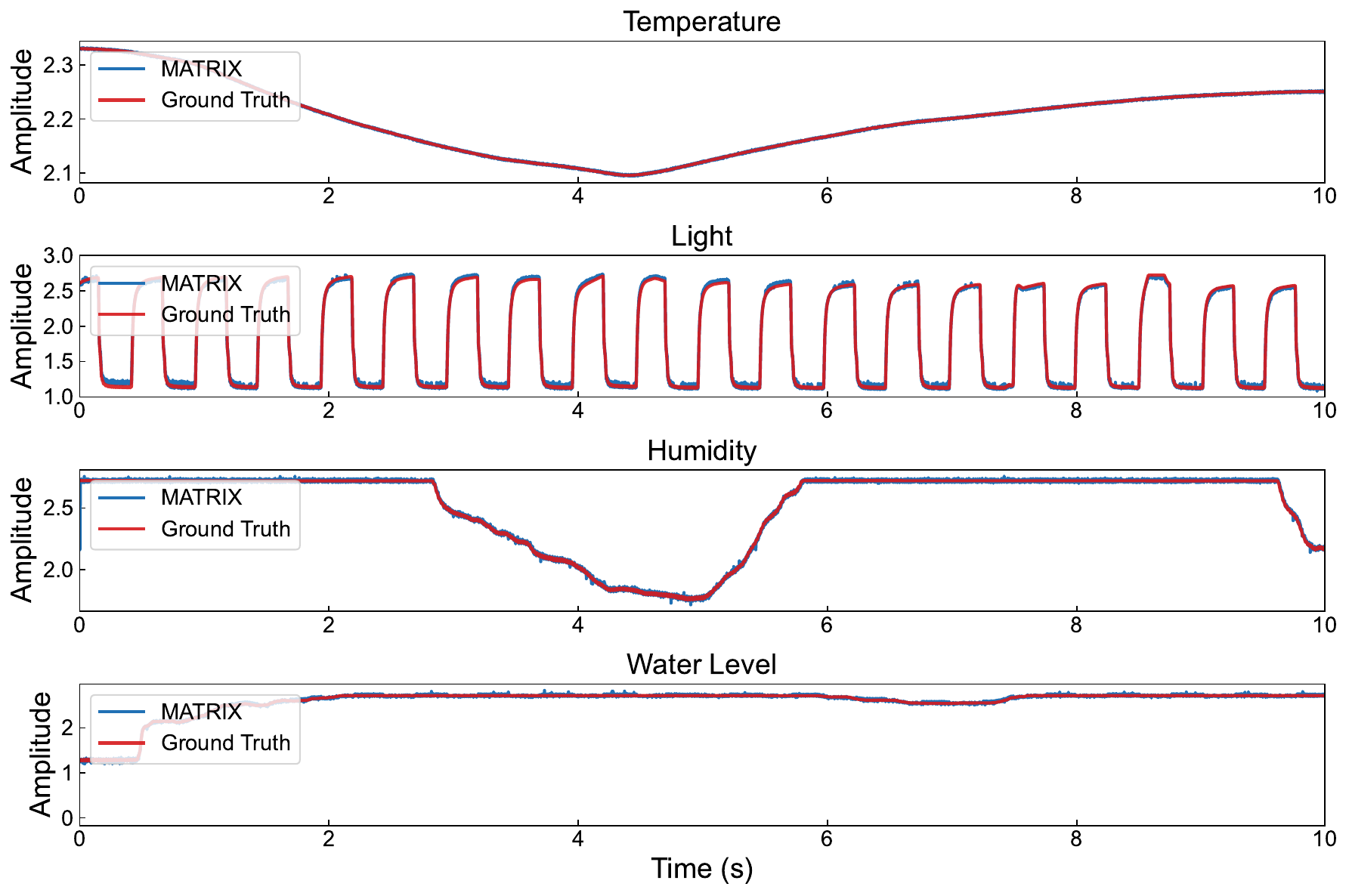}
	\vspace{-12pt}
	\caption{Plant sensing with 4 sensor streams.} 
	\label{fig:plantsensing}
	\vspace{-10pt}
\end{figure}

\sssec{Plant Sensing:} Plant health monitoring systems~\cite{lu2020multimodal} often require multiple types of sensors on a single site. We demonstrate  the \sysname tag for plant sensing by connecting it with off-the-shelf  temperature, optical, soil-moisture,  and  water-level sensors. The TX-to-tag and RX-to-tag distances are  $1m$ and $5m$ respectively. Sensors are sampled at  $1kHz$, which is sufficient for plant sensing. We validate \sysname's functionality by  applying controlled  stimuli: adjusting room temperature, toggling illumination, etc. A synchronous ADC records the ground truth signals, and we reconstruct the sensor readings from the backscatter with \sysname's demultiplexing. Fig.~\ref{fig:plantsensing} shows the ground truth and the reconstructed signals for different sensors. The average SNR is 44 dB, indicating close agreement with the ground truth.

\begin{figure}[t]
	\centering
	\includegraphics[width=0.9\linewidth]{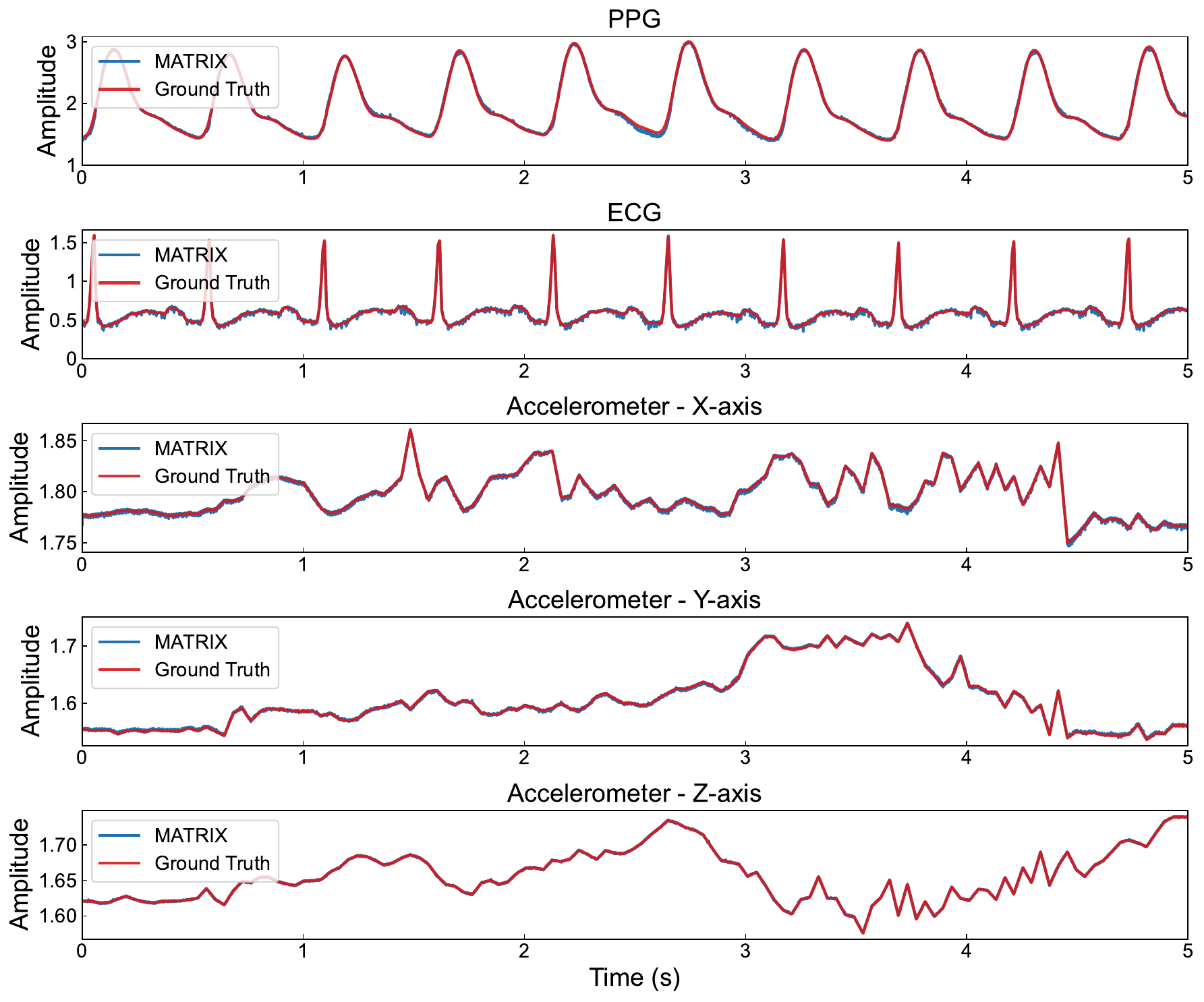}
	\vspace{-15pt}
	\caption{Human health monitoring with 5 sensor streams.} 
	\label{fig:ECG}
	\vspace{-28pt}
\end{figure}

\sssec{Human Health Monitoring:} Human health monitoring often requires the coordinated operation of multiple sensors on a wearable patch, e.g., smart wristband~\cite{park2025ppg}. For example, in cuffless blood pressure monitoring~\cite{laput2017synthetic,shriram2010continuous}, ECG and PPG  are often combined, with accelerometers to suppress motion artifacts. We demonstrate \sysname's capability by integrating with five sensor streams: a PPG sensor, an ECG sensor, and a 3-axis accelerometer (contributing three sensor streams).  All sensors are mounted on the wrist; sensors are sampled at  $1kHz$. A synchronous ADC records  the ground truth. Experiments are conducted  indoors with a TX-to-tag distance of 1 $m$ and an RX-to-tag distance of 10 $m$. During the test, the participant  rotates the wrist to vary the accelerometer readings. Unlike plant sensing, where signals  change slowly, physiological signals exhibit short pulses. \sysname's concurrent acquisition offers advantages in accurately estimating pulse transit time (PTT), which is crucial for blood pressure monitoring~\cite{shriram2010continuous}.  Fig.~\ref{fig:ECG} shows a  5-seconds snapshot of the recoverd signals. We also test  physiological sensors like a respiration sensor and an arterial blood pressure (ABP) sensor. The average SNR is 47dB, showing that \sysname has the potential in health applications that require co-located multi-sensors. 

\begin{figure}
	\centering
	\subfigure[Experiment Setup]{
		\centering
		\includegraphics[width=0.46\linewidth]{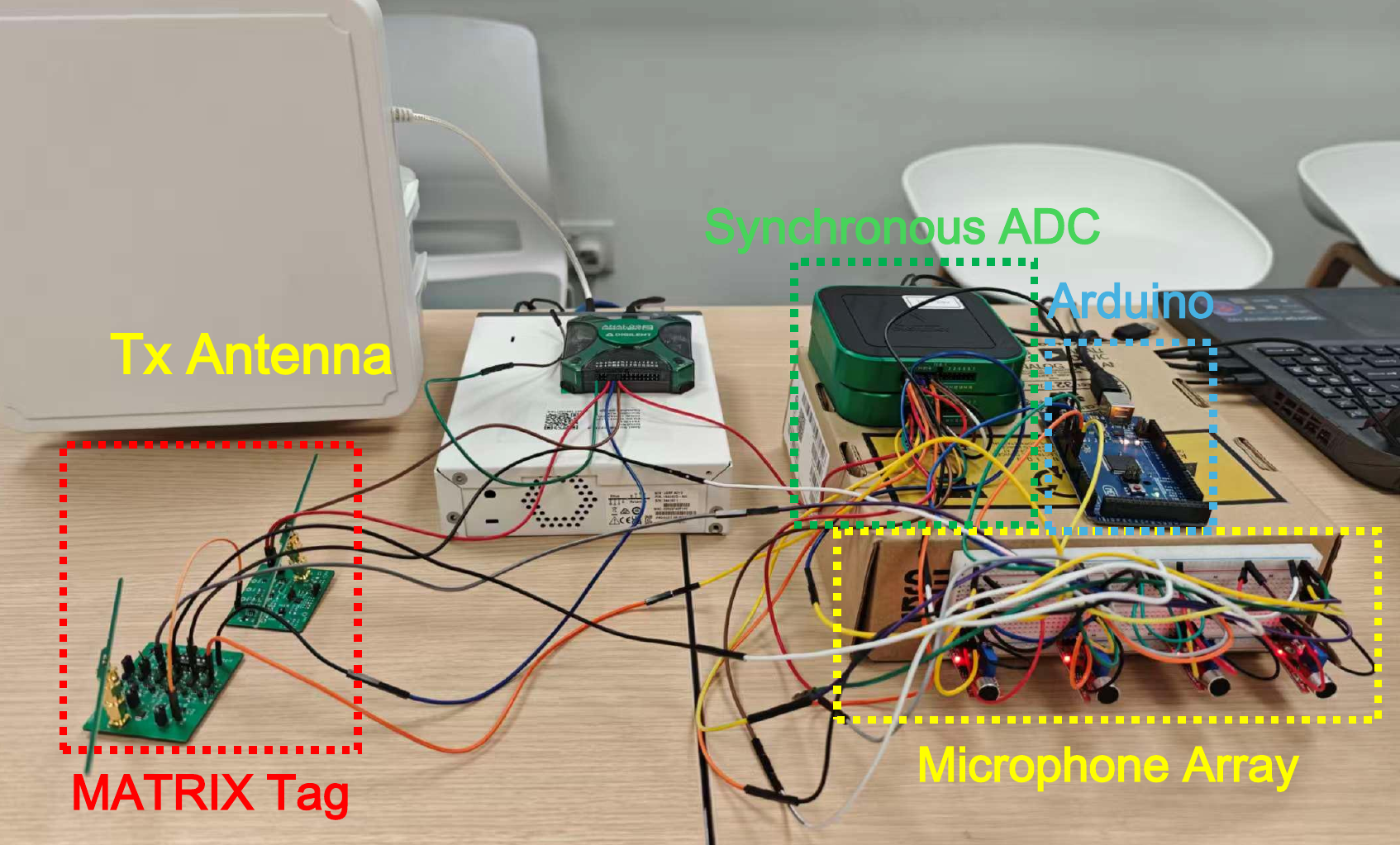}
		\label{fig:AoAsetup}
	}\hspace{10pt}
	\subfigure[AoA estimation error]{
		\centering
		\includegraphics[width=0.45\linewidth]{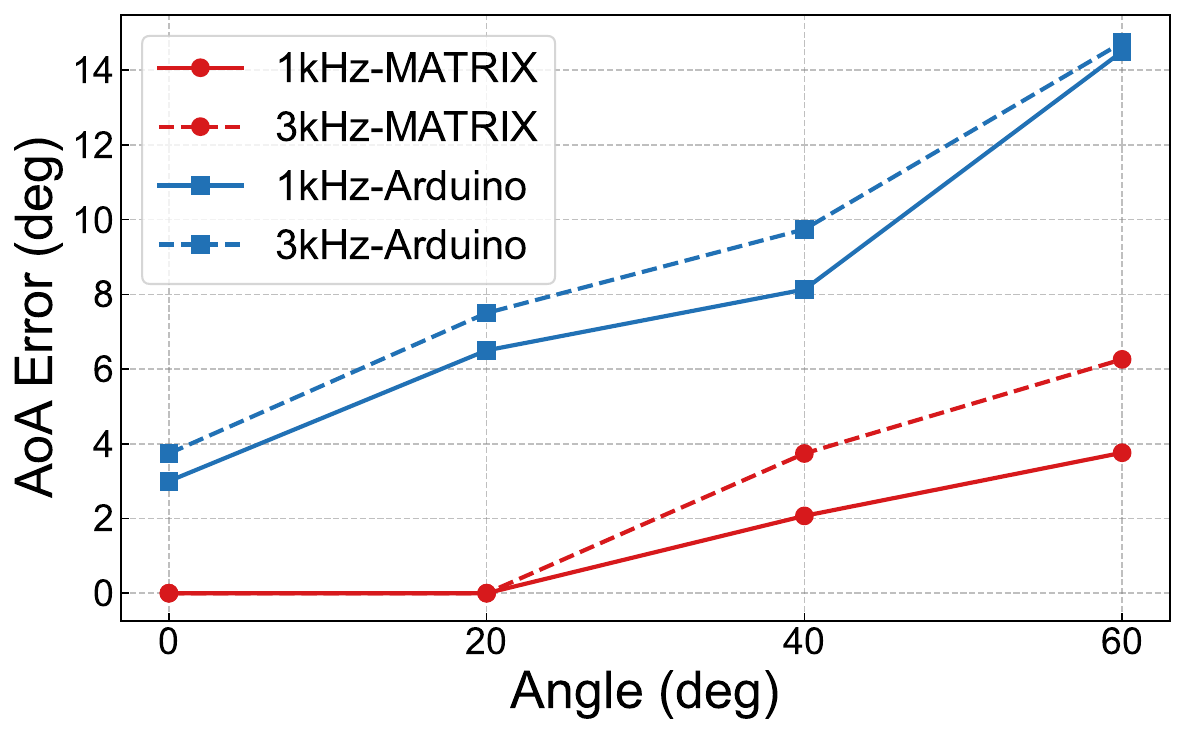}
		\label{fig:AoAresults}
	}
	\vspace{-15pt}
	\caption{AoA test setup. We compare \sysname and a TDM-based Arduino against a synchronous-ADC ground truth.}
	\vspace{-20pt}
\end{figure}

\sssec{Acoustic AoA Estimation:} We test \sysname's capability in acoustic tracking via angle-of-arrival (AoA) estimation with a microphone array. As a baseline, we implement time-division multiplexing (TDM), which sequentially polls sensors under a local clock, introducing inter-sensor sampling offsets; Even a $10\mu s$ offset  leads to $10.8\degree$ of phase errors at a $3kHz$ signal frequency, distorting inter-microphone phase and degrading AoA accuracy. In contrast, \sysname samples all microphones concurrently on a shared time base and multiplexes them for backscatter. To validate \sysname's  effectiveness, we connect four identical microphones to \sysname and compare against (i) an Arduino Mega 2560~\cite{ArduinoMega2560} polling four sensors at 10 kHz (with $13 \mu s$ sampling offsets), and (ii) ground truth: a  4-channel synchronous ADC (power hungry and thus not suitable for a backscatter tag). A single-tone acoustic wave at 1 kHz and 3 kHz is played from different angles using a Sony SBX-100 loudspeaker placed $2m$ away from the microphone array. Higher frequencies are excluded since the Arduino baseline limits four-channel sampling to 12 kHz at 2 M baud rates. All systems thus operate at 10 kHz for fairness. Fig.~\ref{fig:AoAsetup} shows the setup where the microphones'  analog outputs  are sampled in parallel by \sysname, the synchronous ADC (ground truth), and Arduino (TDM baseline).  Fig.~\ref{fig:AoAresults} show that \sysname achieves medium AoA error of $4^\circ$ for the 1 kHz wave and $6^\circ$ for the 3 kHz wave,  outperforming the TDM baseline by $10\degree$ across all tested angles. The gain comes from \sysname's concurrent acquisition as detailed in Sec.~\ref{sec:parallel}, which preserves inter-microphone phase while keeping the tag low-power and single-chain.

%

%


\section{Discussion and Limitations}
\sssec{Hardware Tolerance:} \sysname sets voltage-division weights with high-precision ($\pm 1\%$) resistors to realize accurate placement of the predefined backscatter frequency levels. In practice, the resistor tolerance and hardware aging can cause slight  deviations from the preset frequencies.  \sysname mitigates this issue by using  Gaussian fitting within its HMM to cluster the frequency levels rather than simply calculating absolute values (see Sec.~\ref{sec:HMM}). This removes sophisticated hardware calibration and maintains reliable operation across manufacturing batches and environmental conditions.

\sssec{Number of Sensors Per Tag:} \sysname's prototype concurrently samples five onboard sensors per tag, sufficient for applications such as plant sensing on a single leaf~\cite{lee2023abaxial} or  health monitoring on a wearable patch~\cite{park2025ppg}. Scaling beyond five is  mainly governed by the minimum resolvable frequency spacing, jointly determined by VCO's  frequency  range~\cite{LTC6990} (e.g., $1MHz$) and the receiver's sampling rates ($20MHz$ in our experiments). As more sensors are added, the number of composite levels grows and the inter‑level frequency spacing shrinks within the bounded VCO's bandwidth, making demultiplexing more challenging. To quantify this, we simulate two square-wave signals with  different frequencies and measure the smallest spacing our demodulation can reliably resolve. 
With a 20 $MHz$ sampling rate, \sysname\  distinguishes frequencies separated by $\geq$ 5 $kHz$—sufficient to support seven sensors. Increasing the sampling rate to 100 MHz improves the resolution to 1 kHz, potentially supporting nine sensors, not accounting for added receiver complexity. Balancing tag power and reconstruction accuracy, we evaluate \sysname with five sensors per tag, which already covers the common case of co‑located, multi‑sensor deployments.


\sssec{Multi-tag Scenario: }\sysname's focus is multiplexing multiple onboard sensors on a single tag, not a multiple-access scheme for many tags. However, for scenarios requiring monitoring multiple tags in a shared environment, \sysname can be used together with established anti-collision mechanisms such as RFID's slotted ALOHA or CSMA-based protocols, by incorporating a low-power state machine~\cite{feng2004evaluating} in the tag circuitry, allowing multiple tags to operate in the same environment. Future work could also explore the coordination between tags and the receiver, such as feed-back based collision avoidance~\cite{miskowicz2021unfairness}, further improving scalability in dense tag deployments.

\section{Conclusion}
This paper presents \sysname, a backscatter tag that enables concurrent onboard multi-sensor data acquisition and multiplexing via a single modulation chain.  We develop a voltage-division multiplexing architecture that encodes each sensor reading as a PWM waveform and carefully assigns per‑sensor voltage‑division weights so the resulting composite voltage is well spaced and uniquely invertible back into the original individual sensor readings. The composite voltage is modulated  by an RF switch into backscatter frequency shifts on the carrier. On the receiver, we formulate demultiplexing as a Hidden Markov Model to reliably recover per‑sensor readings despite analog hardware imperfections and multipath.  Our prototype multiplexes five onboard sensors at a 30 kHz sampling frequency and achieves 20 dB average  SNR in signal reconstruction, with demonstrations in health monitoring, plant sensing, and microphone‑based angle‑of‑arrival estimation.


\bibliographystyle{plain}
\bibliography{reference}

\end{document}